\documentclass{pasj00}

\SetRunningHead{Kashiwagi, Yahata \& Suto}
{Detection of FIR Emission from SDSS galaxies}
\Received{2012/07/29}
\Accepted{2012/11/30}
\begin{document}
\title{Detection of Far Infrared Emission from Galaxies and
Quasars \\
in the Galactic Extinction Map by Stacking Analysis}

\author{Toshiya \textsc{Kashiwagi}\altaffilmark{1},
 Kazuhiro \textsc{Yahata}\altaffilmark{1,2}, and Yasushi
\textsc{Suto}\altaffilmark{1,3,4}}
\altaffiltext{1}{Department of Physics, The University of Tokyo,
Tokyo 113-0033}
\altaffiltext{2}{present address: Canon Inc. Ohta-ku, Tokyo 146-8501}
\altaffiltext{3}{Research Center for the Early Universe, School of Science, 
The University of Tokyo, Tokyo 113-0033}
\altaffiltext{4}{Department of Astrophysical Sciences, Princeton
University, Princeton, NJ 08544, USA}
\email{kashiwagi@utap.phys.s.u-tokyo.ac.jp}


\maketitle

\begin{abstract}
We have performed stacking image analyses of galaxies over the Galactic
extinction map constructed by \citet{SFD98}. We select $\sim 10^7$
galaxies in total from the Sloan Digital Sky Survey (SDSS) DR7
photometric catalog. We detect clear signatures of the enhancement of
the extinction in $r$-band, $\Delta A_r$, around galaxies,
indicating that the extinction map is contaminated by their FIR (far
infrared) emission.  The average amplitude of the contamination per
galaxy is well fitted to $\Delta A_r(m_r) = 0.64 \times
10^{0.17(18-m_r)}~{\rm [m mag]}$.  While this value is very small, it is
directly associated with galaxies and may have a systematic effect on
 galaxy statistics.  Indeed this correlated contamination leads to a
relatively large anomaly of galaxy surface number densities against the
SFD extinction $A_{\rm SFD}$ discovered by \citet{Yahata2007}.
We model the radial profiles of stacked galaxy images, and find that the
FIR signal around each galaxy does not originate from the central
galaxy alone, but is dominated by the contributions of nearby galaxies via
galaxy angular clustering. The separation of the single galaxy and the
clustering terms enables us to infer the statistical relation of the FIR
and $r$-band fluxes of galaxies and also to probe the flux-weighted
cross-correlation of galaxies, down to the magnitudes that are difficult
to probe directly for individual objects. We repeat the same stacking
analysis for SDSS DR6 photometric quasars and discovered the similar
signatures but with weaker amplitudes. The implications of the present
results for galaxy and quasar statistics and for correction to the
Galactic extinction map are briefly discussed.
\end{abstract}

\section{Introduction}

Galactic extinction is the most fundamental correction that should be
applied to virtually all astronomical data.  The most widely used map
for the purpose is constructed by Schlegel, Finkbeiner \& Davis (1998;
hereafter, SFD). They first made dust temperature and emissivity maps
($\timeform{0D.7}$ FWHM spatial resolution) from COBE/DIRBE data at
$100~{\mu \rm{m}}$ and $240~{\mu \rm{m}}$. Then a finer resolution map was created
for dust emission ($\timeform{6'.1}$ FWHM angular resolution) from IRAS/ISSA
data at $100~{\mu \rm{m}}$ using the COBE temperature map as a
calibrator. Finally they constructed the maps of reddening and
extinction assuming a simple linear relation between far infrared (FIR) flux at 100$\mu \rm{m}$ and
dust column density with a temperature correction for dust emissivity.
As is clear from their construction procedure, the SFD map {\it does
not} correspond to absorption-weighted, as required for most purposed in
astronomy, but is an emission-weighted dust extinction map.

This motivated Yahata et al.(2007; hereafter Y07) to examine the
validity of the assumed proportionality between extinction and
emissivity using the surface density of SDSS \citep{York2000} galaxies
in a given apparent magnitude range as a calibrator.  Surprisingly, Y07
found that the galaxy surface densities {\it increase} against the value
of the extinction $A_{r,\rm{SFD}}$ for low extinction regions
($A_{r,\rm{SFD}}<0.1$); see their Figure 4.  After several careful
analyses, they concluded that this anomaly originates from the far
infrared (FIR) emission from the SDSS galaxies themselves. If such
additional FIR emissions from extra-galactic objects are incorrectly
interpreted as those from the Galactic dust, the extinction of those
regions is overestimated. The amount of the FIR emission would be an
increasing function of the surface number density of galaxies, and the
correlation between them should be visible especially where the Galactic
extinction is small. Moreover the correlation should become even
stronger after correction using the contaminated extinction value.

Although the proposal of Y07 for the origin of the anomaly is
reasonable, it is admittedly based on indirect and circumstantial
evidence. In the present {\it paper}, we report on a detection of FIR
emission signatures from SDSS galaxies and quasars through a stacking
analysis of the SFD map. The amount of the FIR emission from those
objects is statistically consistent with that required to explain the
anomaly discovered by Y07.

The present paper is organized as follows; in \S 2 we describe part of
the SDSS DR7 data catalog that we use in the analysis. After presenting
the stacking analysis method, we show the stacked images and the
resulting fit to the analytical profile. Section 3 discusses the
physical interpretation of the extracted parameters for FIR emission
from SDSS galaxies and comparison with the cross-correlation with IRAS
PSCz catalog. We also repeat the similar stacking
analysis for SDSS DR6 photometric quasars in Section 4. Section 5
summarizes the conclusions. Finally in Appendix we show the IRAS Point
Spread Function that we measure.

\section{Stacking analysis}

\subsection{SDSS DR7 Data}

In the following analysis, we use the SDSS DR7 photometric galaxy
 catalog, which covers 11663 $\rm{deg^2}$ of sky area, with
 photometry in five passbands; $u$, $g$, $r$, $i$, and $z$ (For more
 details of the photometric data, see \cite{Stoughton2002, Gunn1998,
 Gunn2006, Fukugita1996, Hogg2001, Ivezic2004, Smith2002, Tucker2006,
 Padmanabhan2008, Pier2003}).  According to photometry processing flags,
 we carefully remove bad photometry data and fast-moving objects, which
 are likely in the Solar system or associated with the interplanetary dust.
 We also exclude objects in masked regions to avoid unreliable
 photometry data.  For more details of our data selection, see
 \citet{Yahata2007}.

Our analysis below does not exclude the galaxies that
 are also detected by IRAS PSCz (\cite{Saunders2000}; see \S 3) that are removed in
 the SFD map.  We made sure that this has a negligibly small effect on
 our result because of the small number of overlapped galaxies.

\subsection{Stacking Method}

\begin{figure}[bt]
\begin{center}
    \FigureFile(70mm,70mm){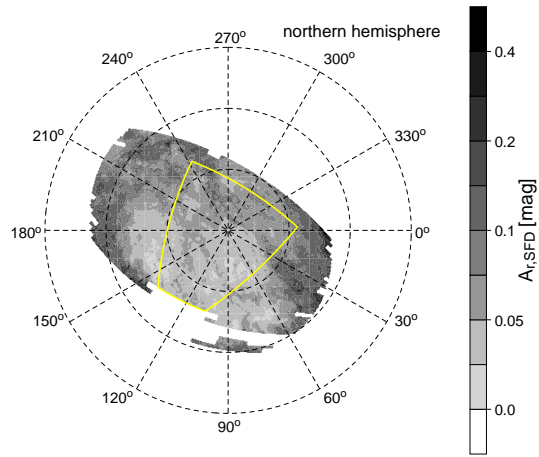}
\end{center}
\caption{The region of the sky used for the present analysis, which is
 shaded according to the extinction value $A_{r,\rm{SFD}}$.
The yellow lines indicate the inner regions used for comparison 
 in Subsection \ref{subsec:galaxies}.}
 \label{fig:region-SFDmap}
\end{figure}

As discussed by Y07, the amount of FIR emission from SDSS galaxies that
explains the anomaly is very small, and it is impossible to detect for
individual galaxies. Therefore we stack those regions of the SFD map
centered at the positions of SDSS photometric galaxies (in this section)
or quasars (in appendix) over their appropriate magnitude bins.
For that purpose, we use a contiguous region in the SDSS DR7 photometric
catalog as shown in Figure \ref{fig:region-SFDmap}.  

\begin{figure}[bt]
\begin{center}
    \FigureFile(70mm,70mm){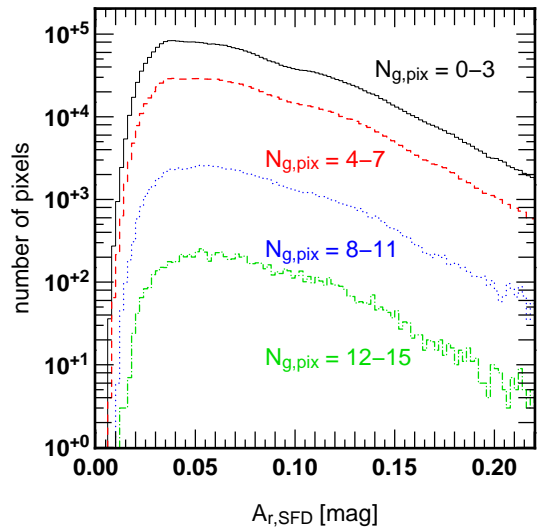}
\end{center}
\caption{Histograms of $A_{r,\rm{SFD}}$ for $\timeform{2'.37} \times \timeform{2'.37}$
 pixels over our selected survey region of SDSS DR7 as a function of the
 number of galaxies within the pixel, $N_{\rm g,pix}$;
$N_{\rm g,pix}=0-3$ in black, $4-7$ in red, $8-11$ in blue, and $12-15$ in
 green.}
 \label{fig:histogram-ASFD}
\end{figure}

\begin{figure}[bt]
\begin{center}
    \FigureFile(70mm,70mm){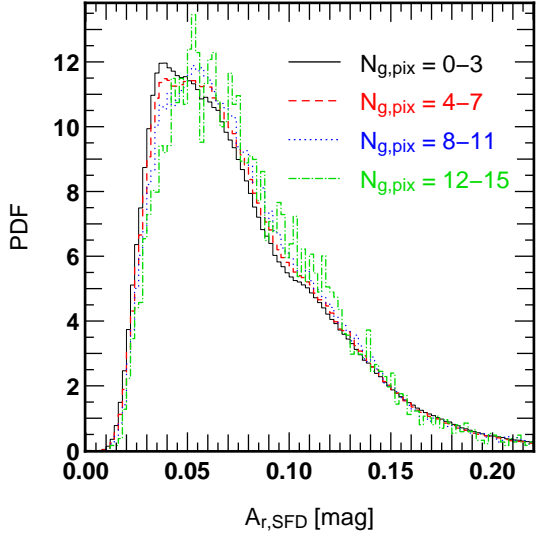}
\end{center}
\caption{The normalized probability density function (PDF) of
 $A_{r,\rm{SFD}}$
corresponding to Figure \ref{fig:histogram-ASFD}.}
 \label{fig:pdf-ASFD}
\end{figure}

The original SFD map divides all sky area into $\timeform{2'.37} \times
\timeform{2'.37}$ pixels and the extinction value is provided for the
central position of each pixel.  The histograms of $A_{r,\rm{SFD}}$
evaluated at those pixels as a function of the number of galaxies with
$15.5<m_r<20.5$ within the pixel, $N_{\rm g,pix}$, are shown in Figure
\ref{fig:histogram-ASFD}.  While the overall shapes of the histograms
are very similar for different $N_{\rm g,pix}$, the
normalized probability density function (PDF) plotted in Figure
\ref{fig:pdf-ASFD} exhibits the small but systematic shift toward the
larger $A_{r,\rm{SFD}}$ with increasing $N_{\rm g,pix}$.  This indicates
the correlation of the Galactic extinction and the background galaxies
that will be extensively discussed in the present paper.

\begin{figure*}[bt]
\begin{center}
    \FigureFile(45mm,45mm){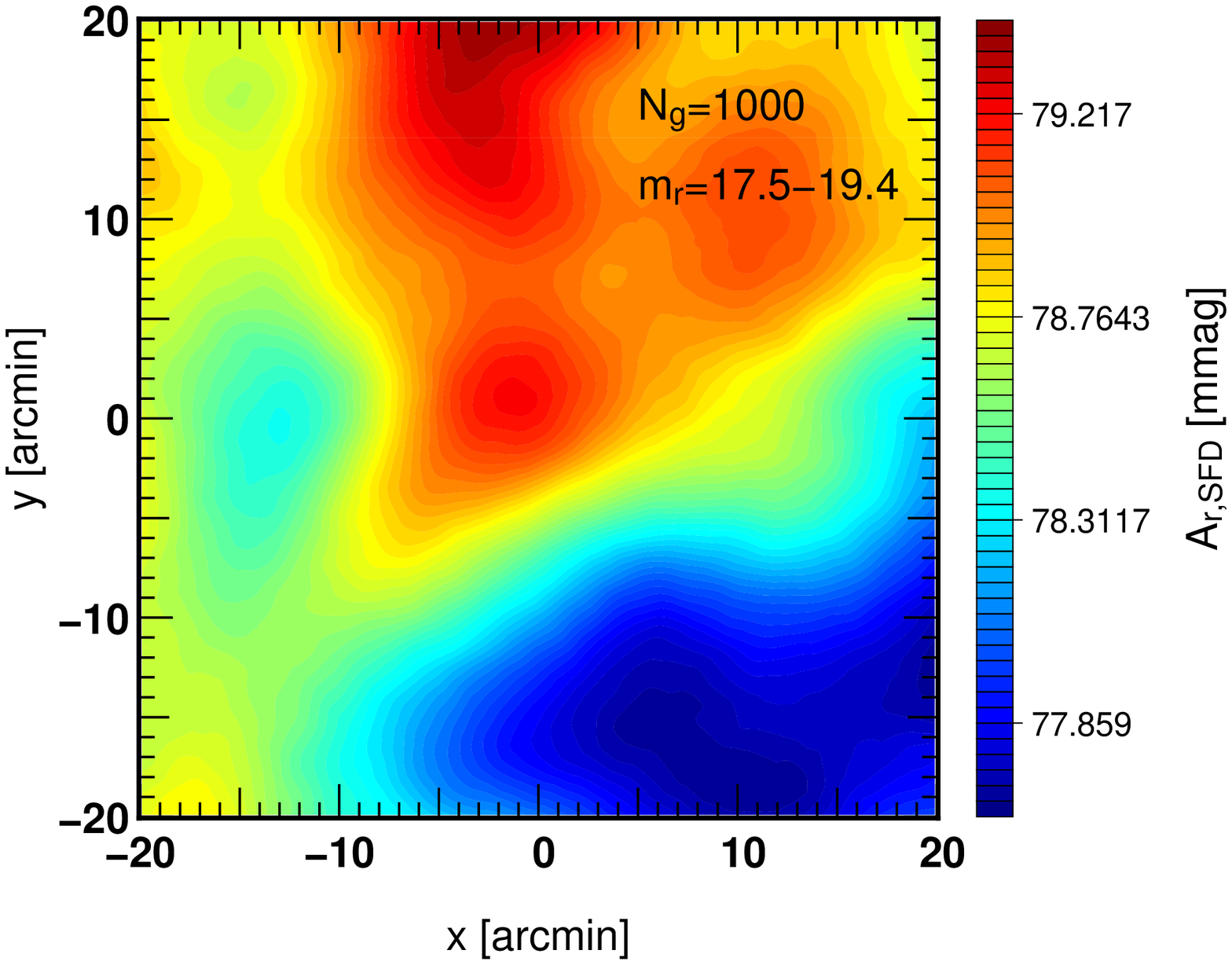}
    \FigureFile(45mm,45mm){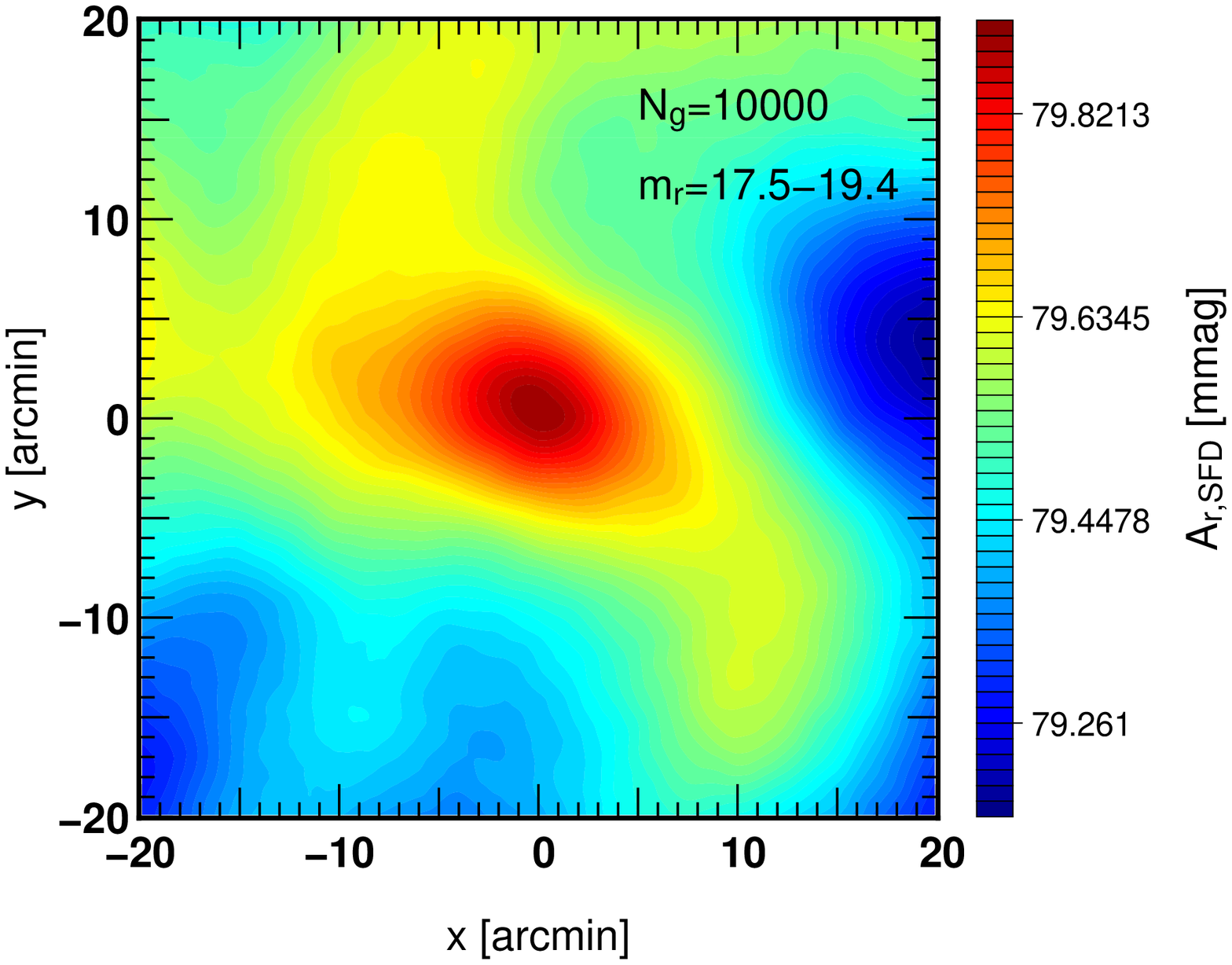}
    \FigureFile(45mm,45mm){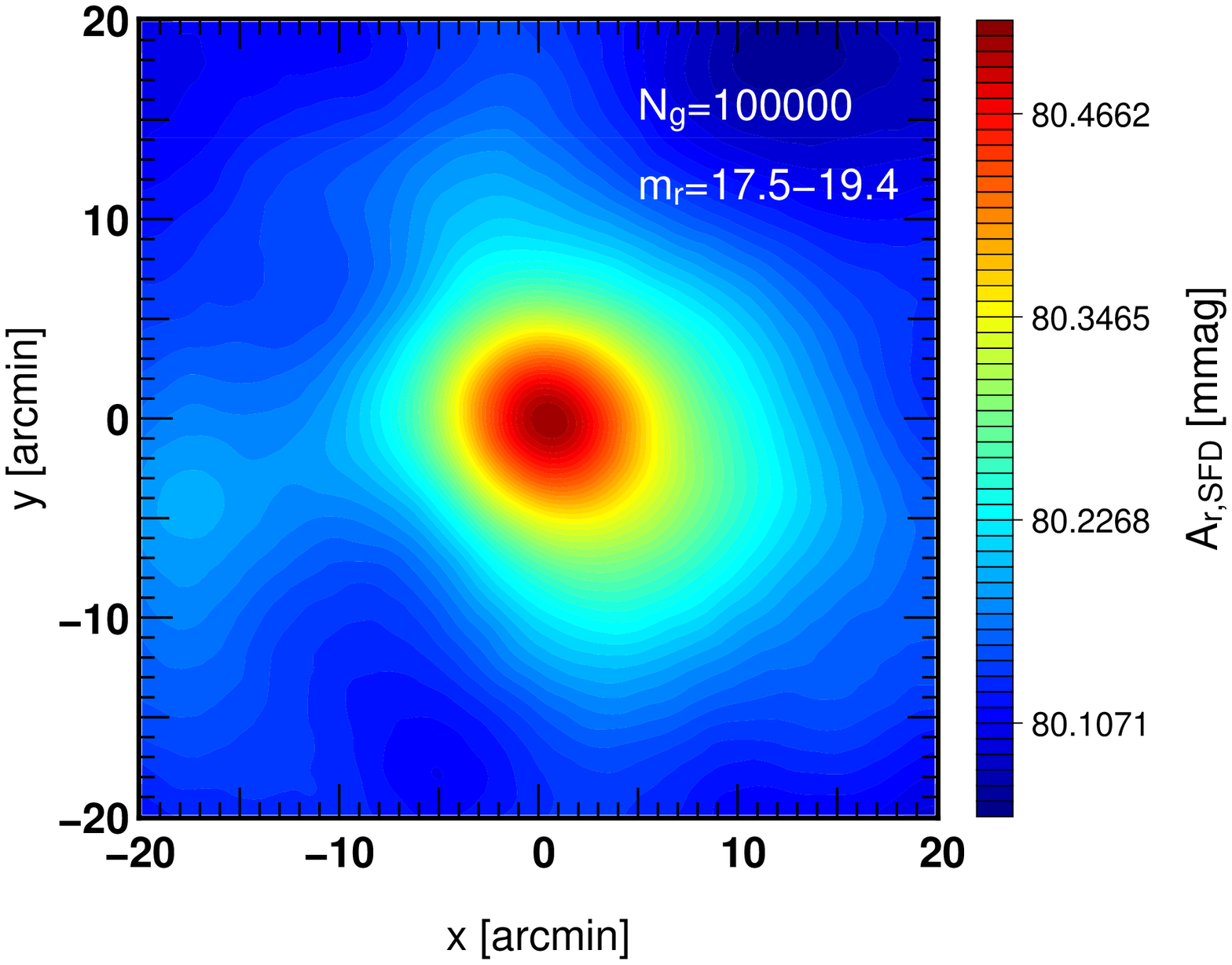}
    \FigureFile(45mm,45mm){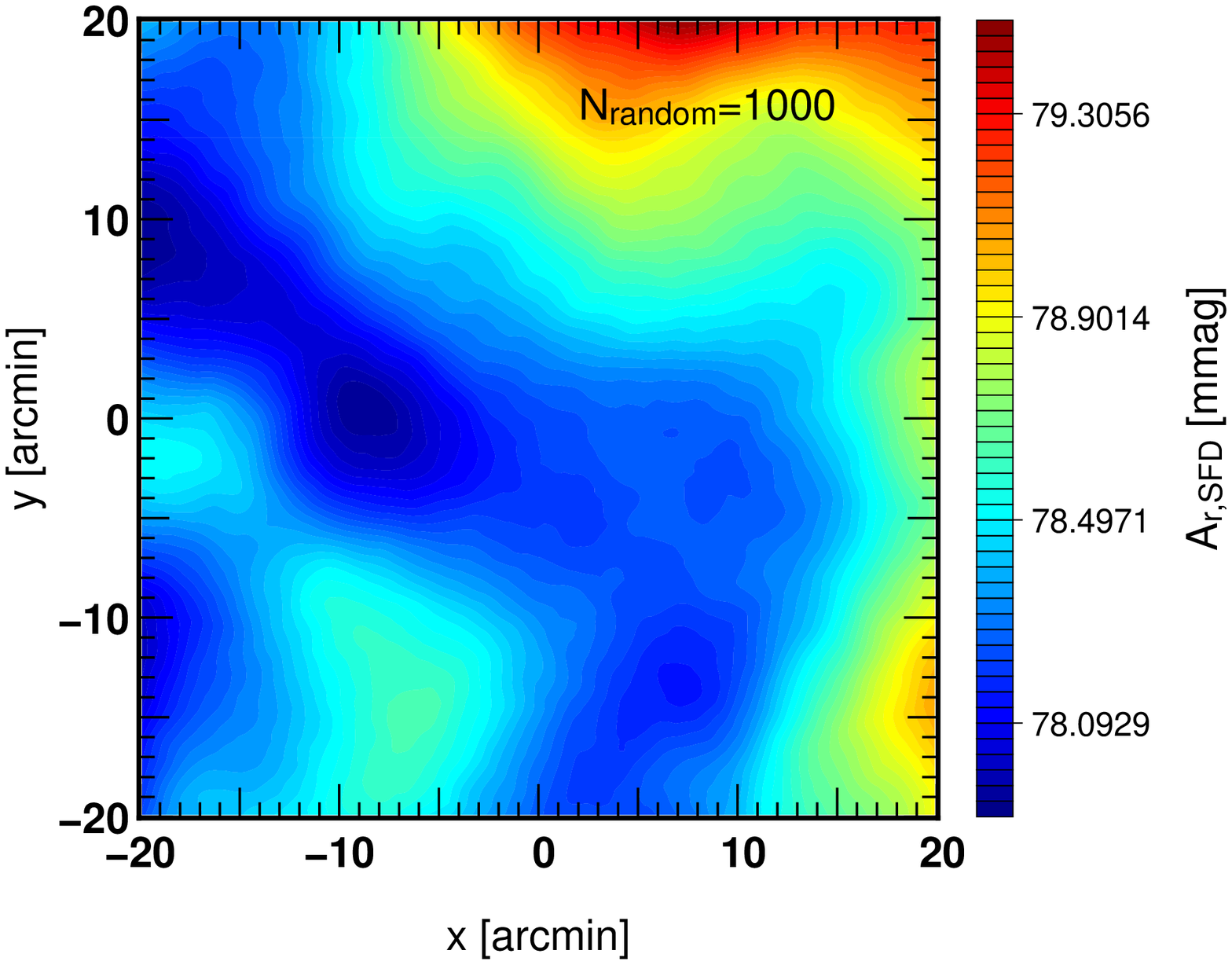}
    \FigureFile(45mm,45mm){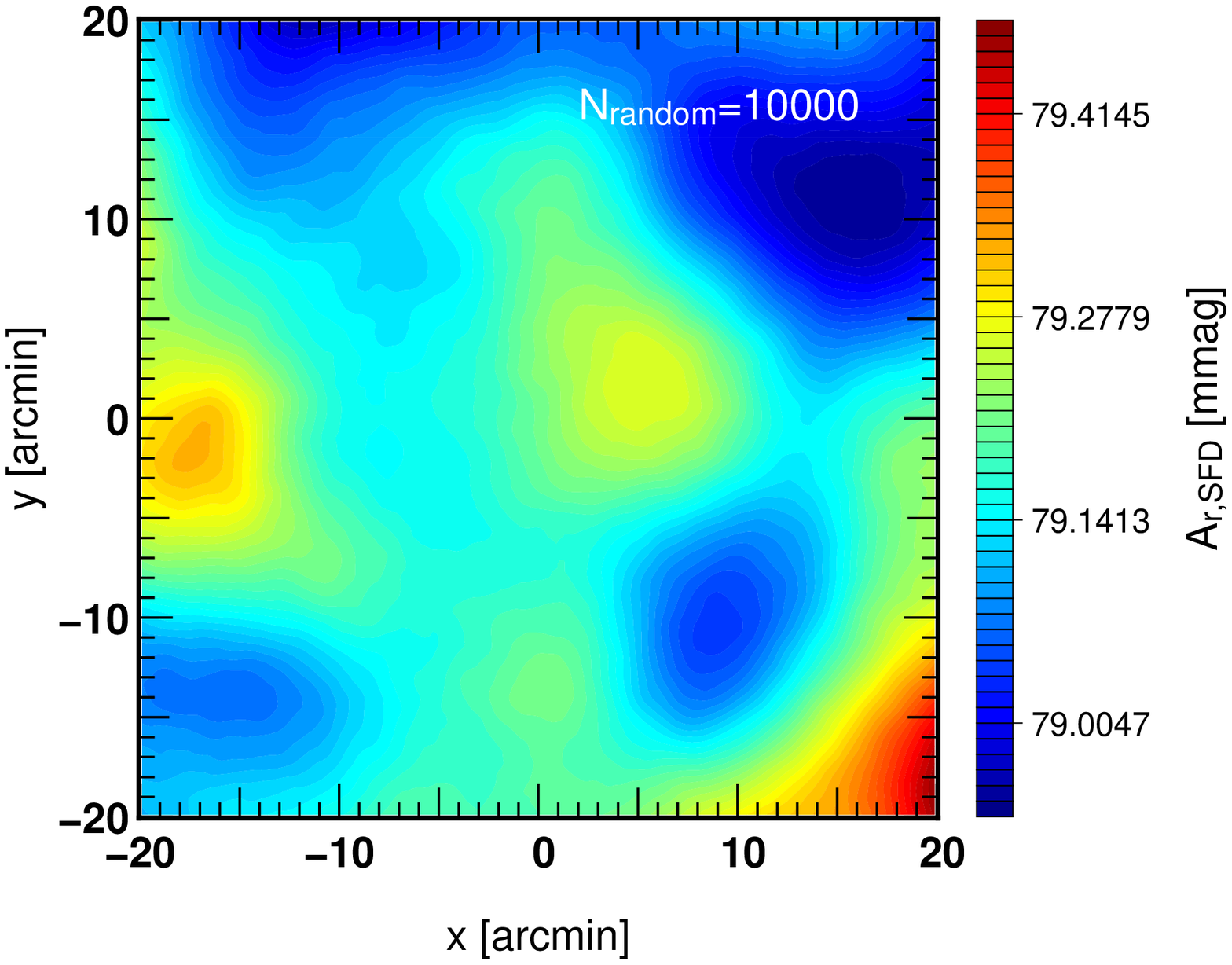}
    \FigureFile(45mm,45mm){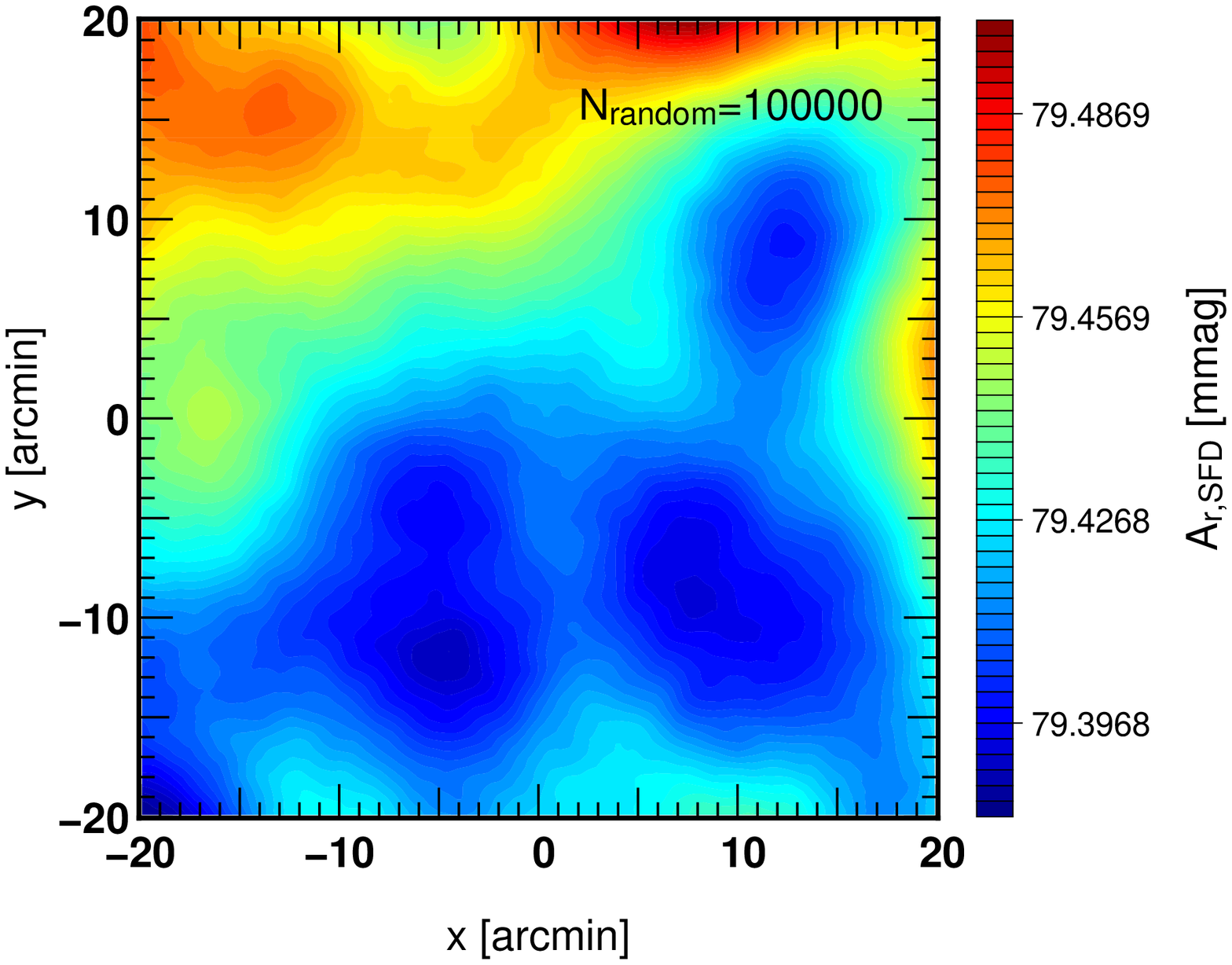}
\end{center}
\caption{Stacked images of the SFD map for $40' \times 40'$ regions;
{\it Upper panels} are centered at the positions of SDSS galaxies of
$17.5<m_r<19.4$, and {\it lower panels} show the reference images
centered at randomly selected positions.
Left, middle and right panels correspond to images stacking $10^3, 10^4, 10^5$ images.
}
\label{fig:stacked-random-images}
\end{figure*}

First we show the result of stacked SFD map images centered at
photometric galaxies in the $r$-band magnitude range of $17.5<m_r<19.4$
randomly selected from a contiguous region in Figure
\ref{fig:region-SFDmap}. In this procedure we evaluate the value of
$A_{\rm r,SFD}$ on $\timeform{0'.2} \times \timeform{0'.2}$ pixels over 
$\timeform{40'}\times \timeform{40'}$ images by
cloud-in-cell interpolation of the 4 nearest neighbors in the original
SFD pixels. Upper panels of Figure \ref{fig:stacked-random-images}
clearly show the presence of the strong feature of $A_{r,\rm SFD}$
around SDSS galaxies, which becomes more pronounced as increasing the
number of stacked galaxies. For reference, lower panels of Figure
\ref{fig:stacked-random-images} show the stacked SFD map images centered
at the same number of randomly chosen positions from the same region of
the corresponding top panels.  This result directly confirms the
interpretation of Y07 that the SFD map is contaminated by the FIR
emission from SDSS galaxies.

\subsection{Radial profiles of galaxies \label{subsec:galaxies}}

\begin{figure*}[t]
\begin{center}
    \FigureFile(45mm,45mm){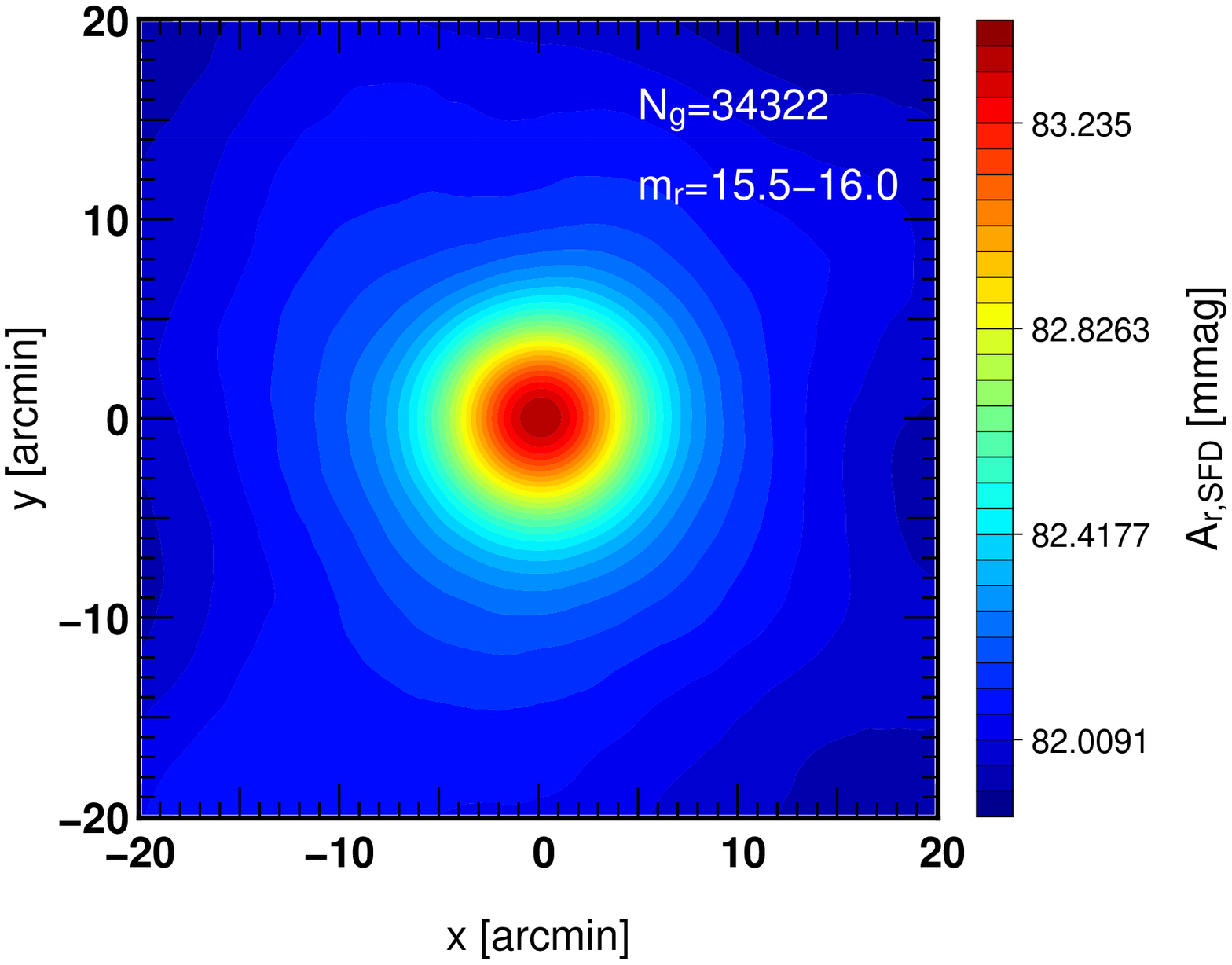}
    \FigureFile(45mm,45mm){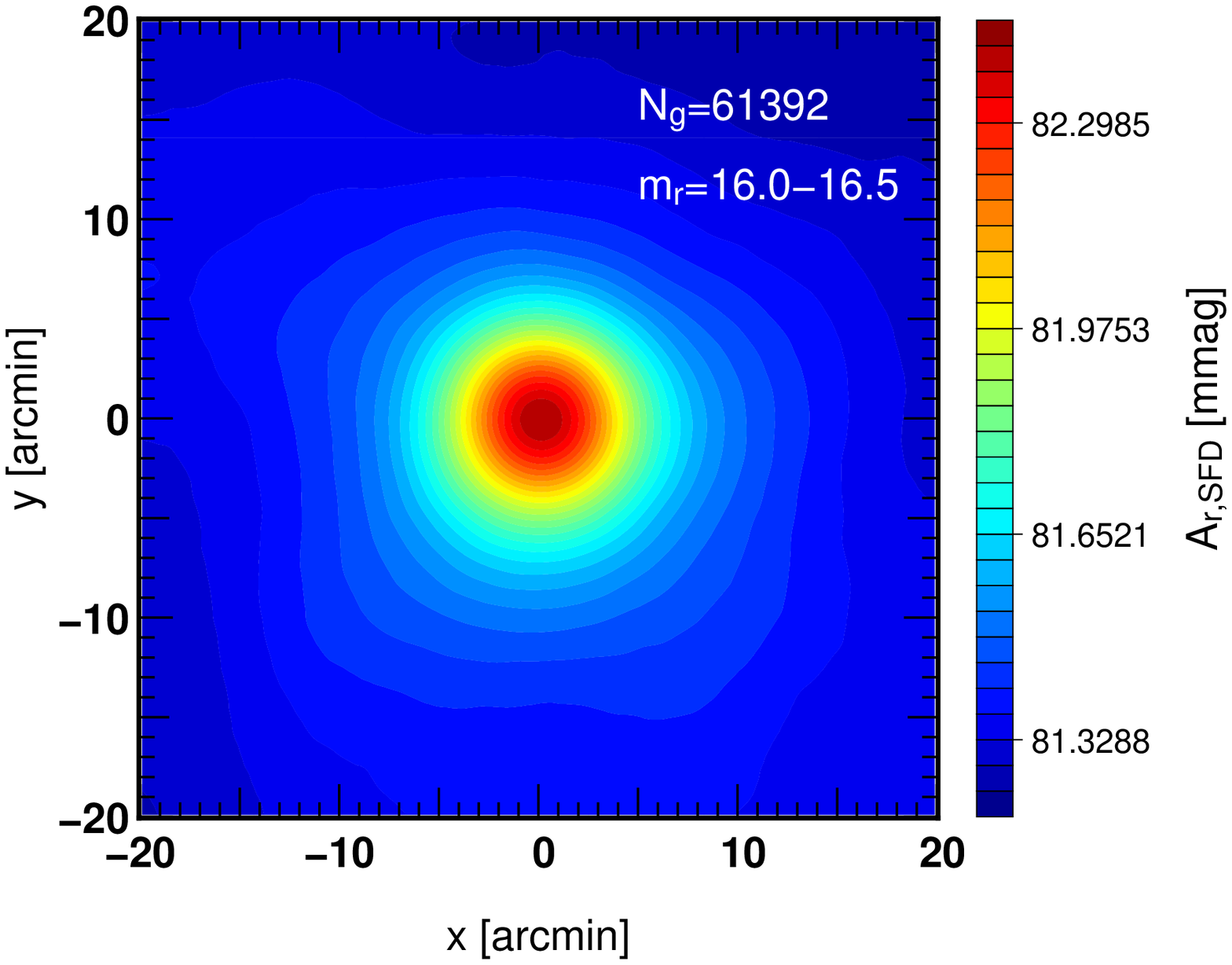}
    \FigureFile(45mm,45mm){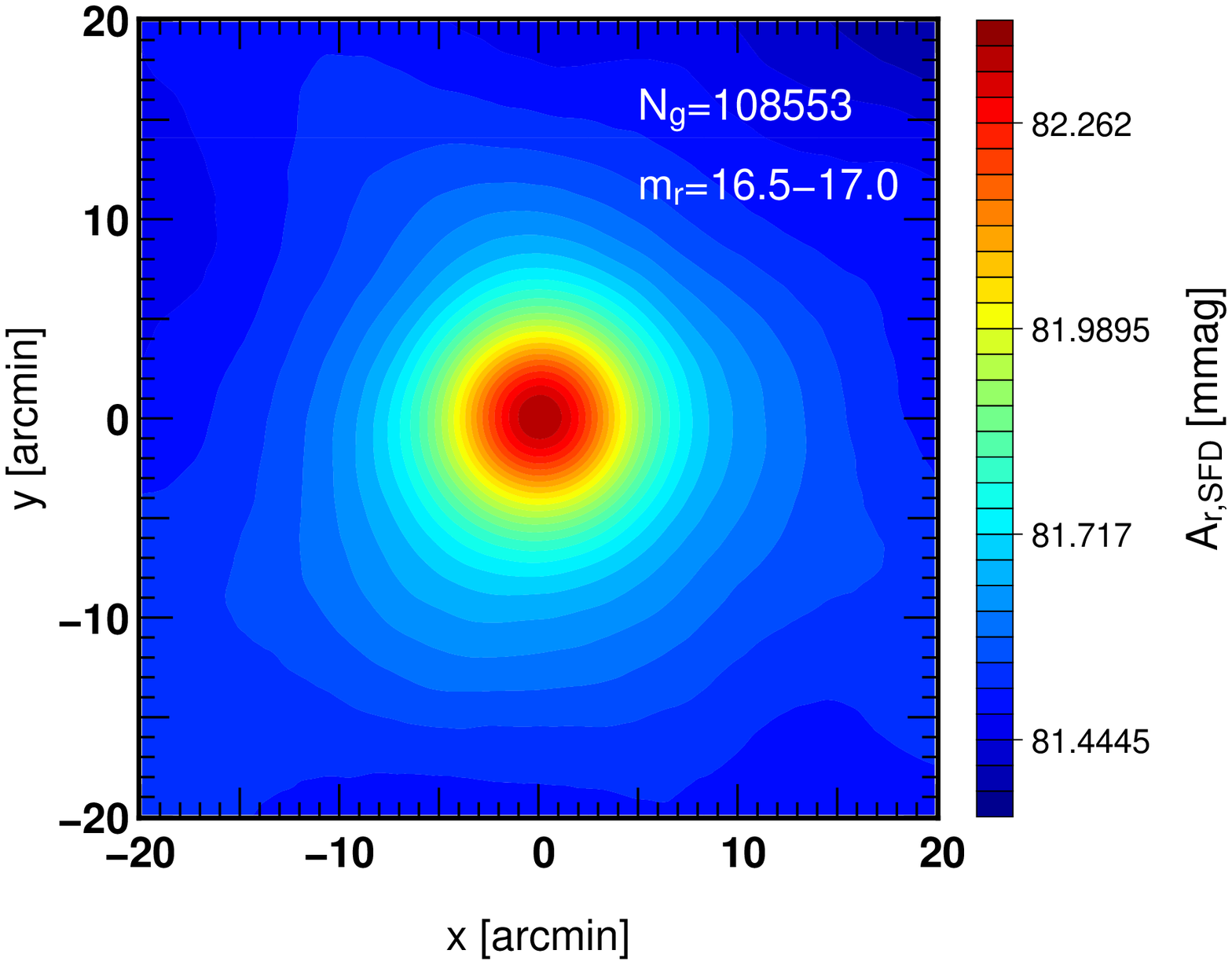}
    \FigureFile(45mm,45mm){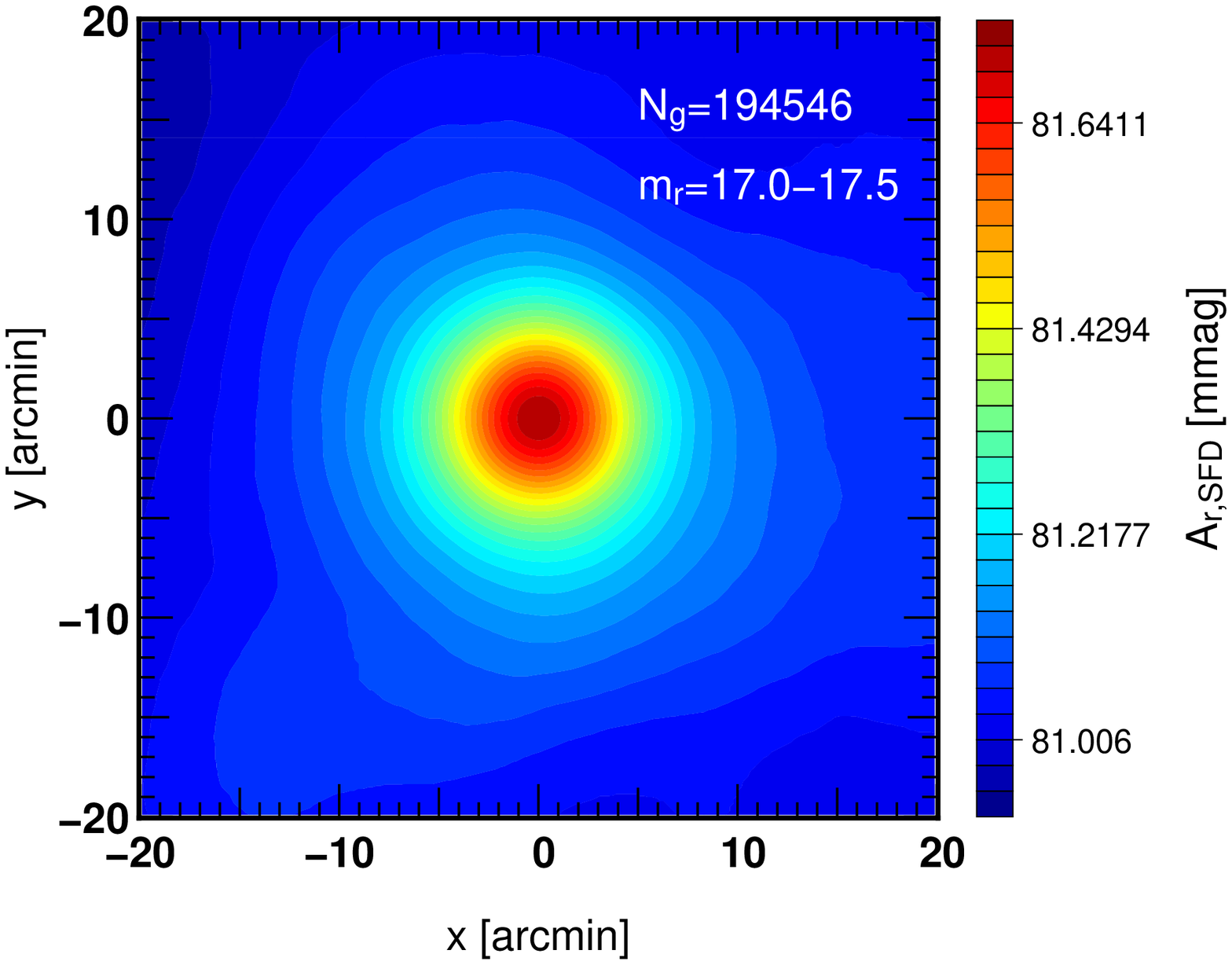}
    \FigureFile(45mm,45mm){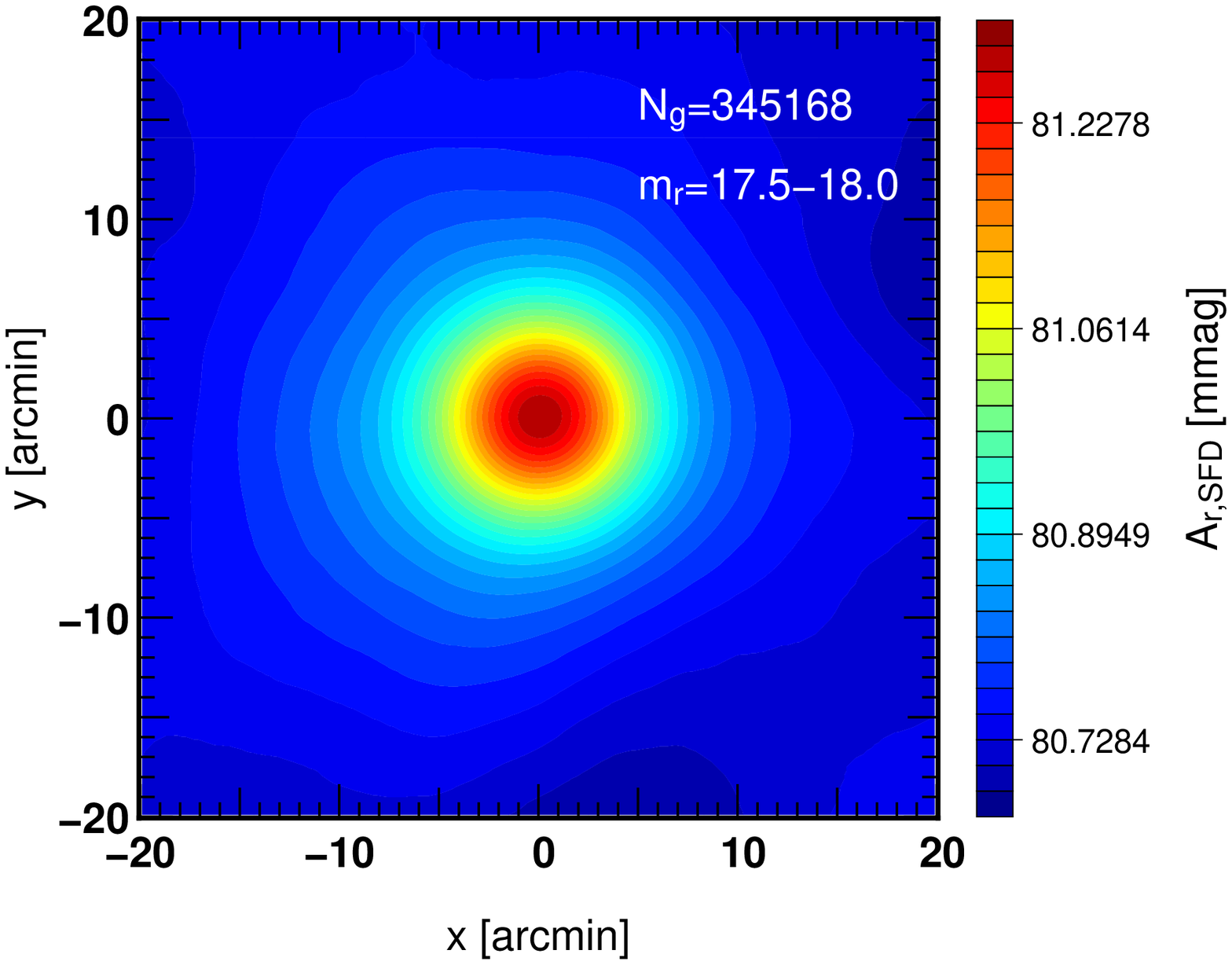}
    \FigureFile(45mm,45mm){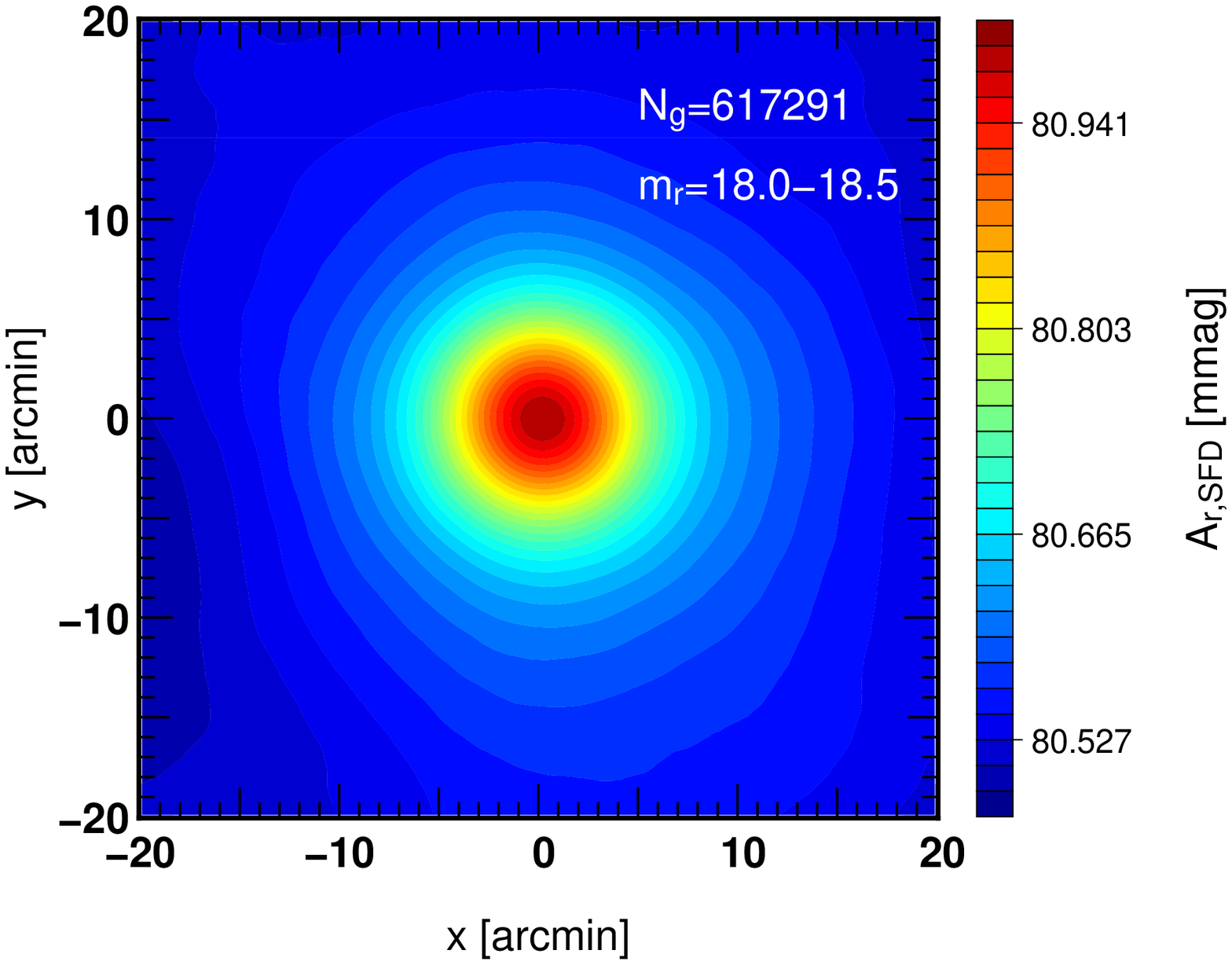}
    \FigureFile(45mm,45mm){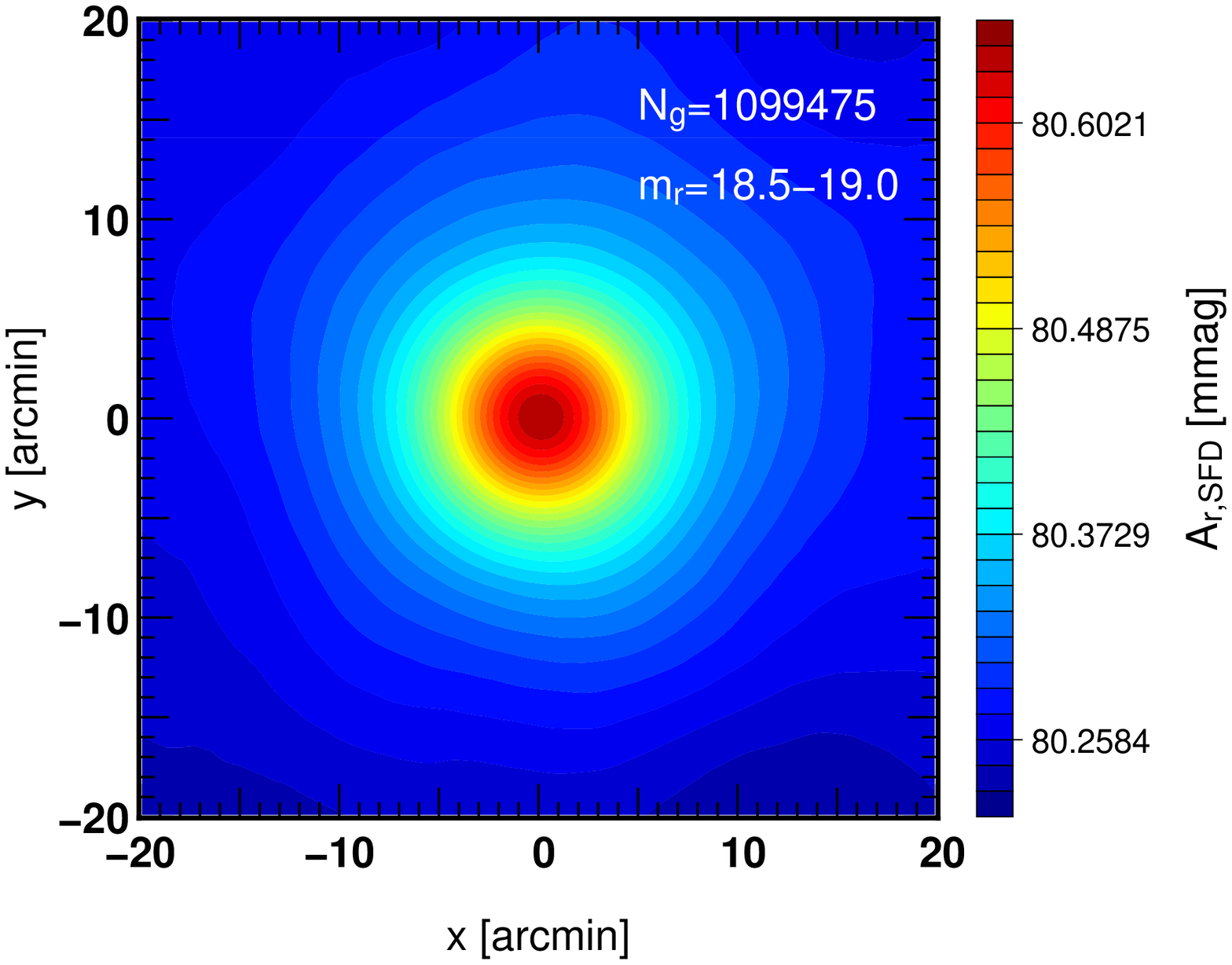}
    \FigureFile(45mm,45mm){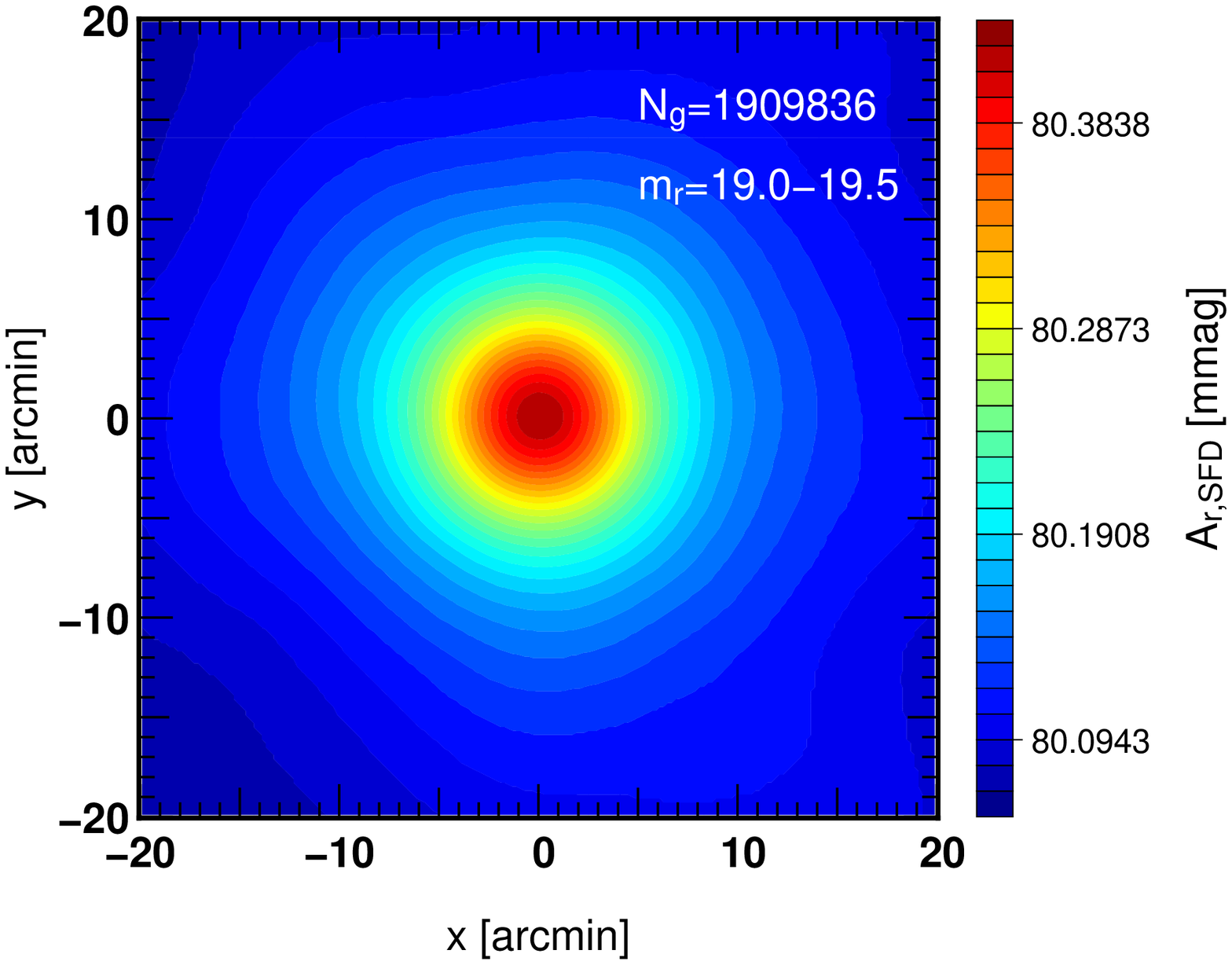}
    \FigureFile(45mm,45mm){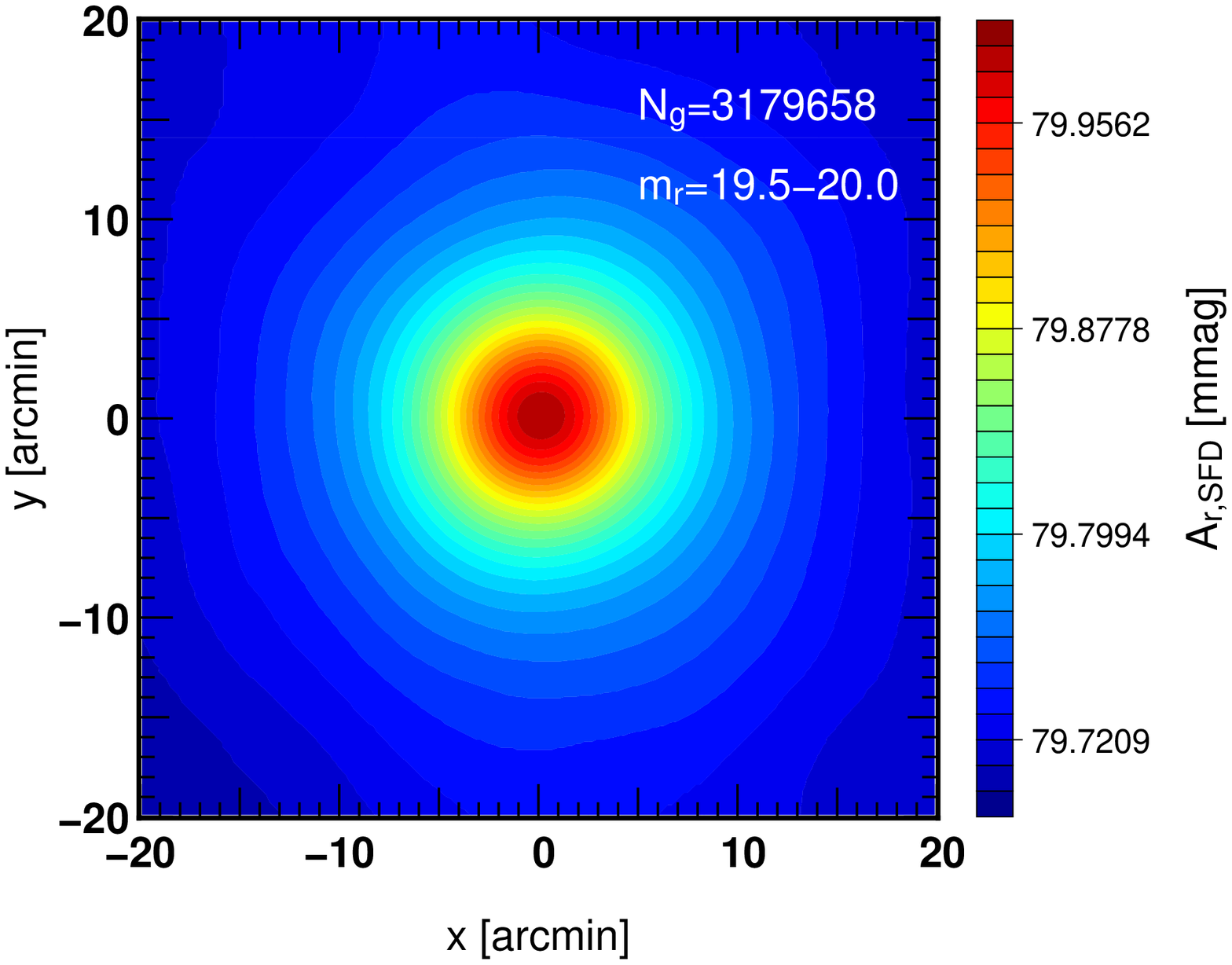}
    \FigureFile(45mm,45mm){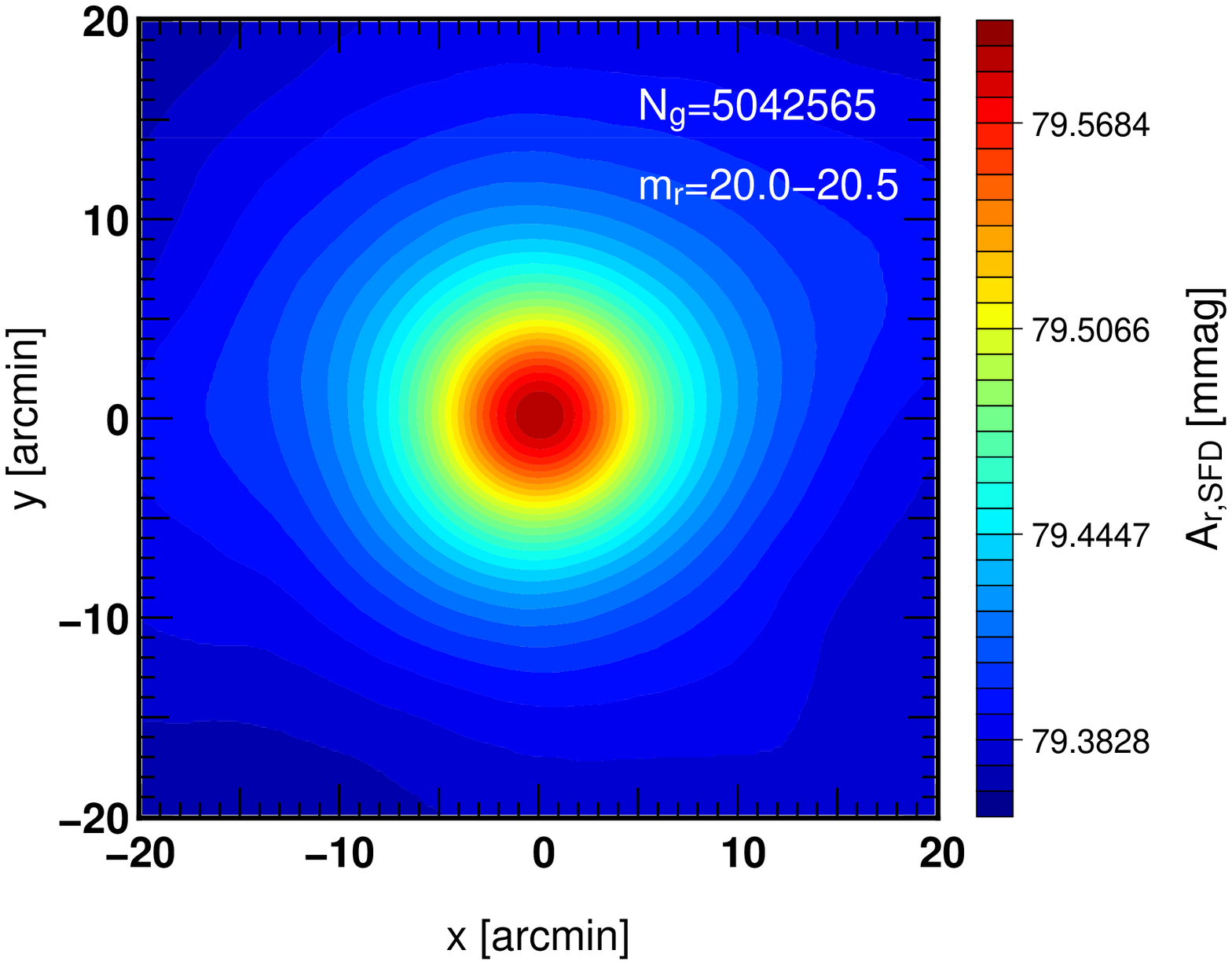}
\end{center}
\caption{Stacked images of the SFD map for $40'\times40'$ centered at
SDSS galaxies of different $r$-band magnitudes ($m_r=15.5\sim20.5$mag.) 
in 0.5 magnitude bin. The magnitude range and the number of galaxies in 
the range are denoted in each panel.}  \label{fig:galaxy-diff-mag}
\end{figure*}

To estimate the dependence of the contribution to $A_{r,\rm SFD}$, or
equivalently the amount of the FIR emission, on galaxy $r$-band
magnitudes, we stack the images at the location of galaxies according to
their $r$-band magnitudes. The results are plotted in Figure
\ref{fig:galaxy-diff-mag}. Thanks to the significantly
large number of the SDSS galaxies,
those images are highly circular, assuring
that the signals do not originate from the Galactic contamination.  All
the angular radii of the images are very similar to the expected
smoothing length of the SFD map ($=\timeform{6'.1}$ FWHM), and one may naively
interpret that the central signal is dominated by the single galaxy
contribution. As we will show below, however, this is not the case; in most cases
the signal is rather dominated by the contribution from the nearby
galaxies. Qualitatively this is understood from Figure
\ref{fig:pdf-ASFD} that shows the systematic increase of $A_{r,\rm SFD}$
as a function of numbers of galaxies in the pixel whose size is much
smaller than the overall smoothing size of the SFD map.

\begin{figure}[bt]
\begin{center}
    \FigureFile(70mm,70mm){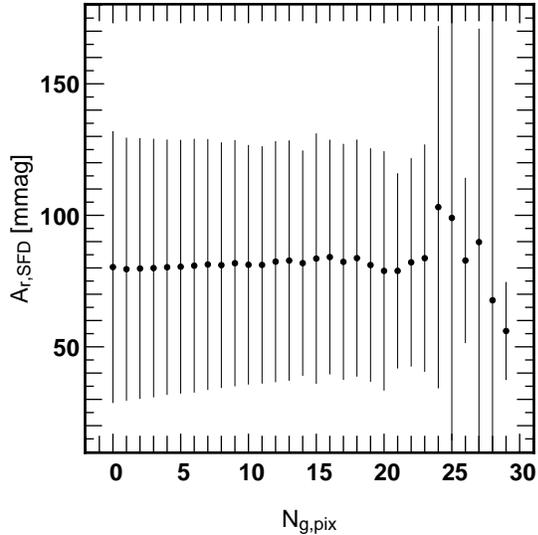}
\end{center}
\caption{Mean values of $A_{r,\rm{SFD}}$ of pixels containing $N_{\rm
g,pix}$ plotted against $N_{\rm g,pix}$. The quoted error bars indicate
the corresponding rms.}  \label{fig:mean-ASFD}
\end{figure}

To proceed more quantitatively, we attempt to model the radial profile of
the stacked images as follows. Denote the angular profile of a single
galaxy with $r$-band magnitude $m_r$ as $\Sigma_{\rm g}^{\rm
s}(\boldsymbol{\theta};m_r)$. Then the statistically averaged
profile of the stacked images centered at the galaxy is given by
\begin{eqnarray}
\label{eq:average-galaxy-profile}
\Sigma_{\rm g}^{\rm tot}({\boldsymbol \theta};m_r) =
\Sigma_{\rm g}^{\rm s}({\boldsymbol \theta};m_r) 
+\Sigma_{\rm g}^{\rm c}({\boldsymbol \theta};m_r)
+ C ,
\end{eqnarray}
where $\Sigma_{\rm g}^{\rm c}(\boldsymbol \theta;m_r)$ denotes the
clustering term corresponding to the contribution from the nearby
galaxies, and $C$ represents the background level of the extinction.
Naively, $C$ is expected to be independent of $m_r$ and
computed from the PDF of the extinction $P(A)$ (see Figure
\ref{fig:pdf-ASFD}) as
\begin{eqnarray}
\label{eq:c-background}
C= \langle A\rangle \pm \frac{\sigma_A}{\sqrt{N_{\rm g}}}, 
\end{eqnarray}
where $N_{\rm g}$ is the number of stacked galaxy images, and the mean
and rms are given by
\begin{eqnarray}
\langle A\rangle &=& \int_0^\infty AP(A)dA, \\
\sigma_A^2 &=& \int_0^\infty A^2P(A)dA - \langle A\rangle^2 .
\end{eqnarray}
As we see below, however, this is not the case.
Therefore we treat $C$ as a free parameter for each magnitude bin in the
fitting analysis described below.  Figure \ref{fig:mean-ASFD} plots
$\langle A\rangle$ with quoted error-bars of $\sigma_A$ as a function of
$N_{\rm g,pix}$.

The clustering term is written as
\begin{eqnarray}
\label{eq:clustering-galaxy-profile}
\Sigma_{\rm g}^{\rm c}({\boldsymbol \theta};m_r) &=&
\iint dm' d{\boldsymbol \varphi}
~\Sigma_{\rm g}^{\rm s}({\boldsymbol \theta-\boldsymbol \varphi};m') \cr
&{}& {~~}\times
w_{\rm g}({\boldsymbol \varphi};m',m_r)
\frac{dN_{\rm g}(m')}{dm'} ,
\end{eqnarray}
where $w_{\rm g}({\boldsymbol \varphi};m',m_r)$ is the angular galaxy
cross-correlation function between magnitudes $m'$ and $m_r$, and 
$dN_{\rm g}(m')/dm'$ is the differential galaxy number density.

Given the large smoothing length of the SFD map ($\timeform{6'.1}$ FWHM), a single
galaxy profile is expected to be approximated by the circular Point
Spread Function (PSF), independently of its intrinsic profile.  Thus we
adopt the Gaussian PSF profile:
\begin{eqnarray}
\label{eq:single-galaxy-profile}
\Sigma_{\rm g}^{\rm s}({\boldsymbol \theta};m_r) 
=\Sigma_{\rm g}^{\rm s0}(m_r)\exp\left(-\frac{\theta^2}{2\sigma^2}\right),
\end{eqnarray}
where $\sigma$ is the Gaussian width of the PSF.  The Gaussian
approximation of the PSF is justified in Appendix. Also we assume that
the angular cross-correlation function is given as
\begin{eqnarray}
\label{eq:wg}
w_{\rm g}({\boldsymbol\varphi};m',m_r)
=K(m',m_r) (\varphi/\varphi_0)^{-\gamma},
\end{eqnarray}
where the constants $\varphi_0$ and $\gamma$ are assumed to be
independent of $m'$ and $m_r$.  We adopt $\gamma = 0.75$
\citep{Connolly2002, Scranton2002}, which is valid for $\varphi<1^\circ$.
With equations (\ref{eq:single-galaxy-profile}) and (\ref{eq:wg}) ,
equation (\ref{eq:clustering-galaxy-profile}) reduces to
\begin{eqnarray}
\label{eq:clustering-galaxy-profile2}
\Sigma_{\rm g}^{\rm c}({\boldsymbol \theta};m_r) &=&
\Sigma_{\rm g}^{\rm c0}(m_r)\exp\left(-\frac{\theta^2}{2\sigma^2}\right) \nonumber \\
&{}& {~~}\times {}_1F_1\left(1-\frac{\gamma}{2};1;\frac{\theta^2}{2\sigma^2}\right), 
\end{eqnarray}
where ${}_1F_1(\alpha;\beta;x)$ is the confluent hypergeometric function, and 
\begin{eqnarray}
\label{eq:sigmac0}
\Sigma_{\rm g}^{\rm c0}(m_r) &=& 
2\pi\sigma^2\left(\frac{\varphi_0}{\sqrt{2}\sigma}\right)^{\gamma}
\Gamma\left(1-\frac{\gamma}{2}\right) \nonumber \\
&{}& {~~}\times \int dm' \Sigma_{\rm g}^{\rm s0}(m')
K(m',m_r) \frac{dN_{\rm g}(m')}{dm'}.
\end{eqnarray}
Equation (\ref{eq:clustering-galaxy-profile2}) results in the extended
tail due to the clustering term in addition to the Gaussian tail of the
single central galaxy. The latter is negligible at $\theta \gg \sigma$,
and the observed tail of the profile around galaxies is basically
dominated by the clustering term.

\begin{figure}[t]
\begin{center}
  \FigureFile(70mm,160mm){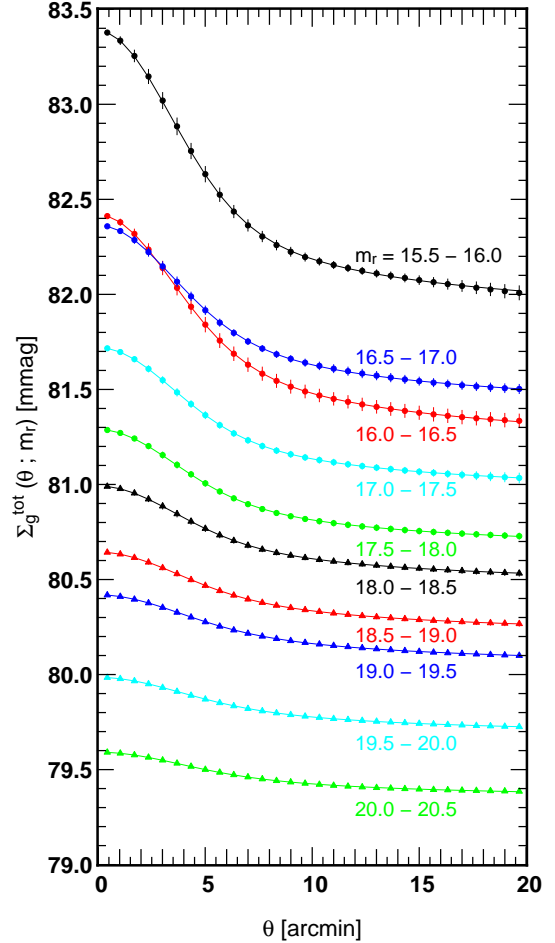}
\end{center}
\caption{Radial profiles of stacked galaxy images corresponding to
Figure \ref{fig:galaxy-diff-mag}. Solid curves indicate
the best-fit model of equation (\ref{eq:average-galaxy-profile}),
(\ref{eq:single-galaxy-profile}), and
(\ref{eq:clustering-galaxy-profile2}).} 
\label{fig:galaxy-profile}
\end{figure}

The average radial profiles of the stacked images centered at
photometric galaxies are plotted in Figure
\ref{fig:galaxy-profile}. Filled circles and triangles
correspond to galaxies in the different $r$-band magnitude ranges in
Figure \ref{fig:galaxy-diff-mag}, and the quoted error-bars represent
rms in each circular bin of $\Delta\theta=\timeform{0'.33}$.  The solid
curves are the best-fit model of equations
(\ref{eq:average-galaxy-profile}), (\ref{eq:single-galaxy-profile}), and
(\ref{eq:clustering-galaxy-profile2}). In fitting
those curves, we treat $\Sigma_{\rm g}^{\rm s0}$, $\Sigma_{\rm g}^{\rm
c0}$, and $C$ for each magnitude bin, and the width of PSF, $\sigma$ as
free fitting parameters.  Here we assume that $\sigma$ is independent
of $m_r$, and fit it simultaneously for all profiles with different
magnitudes. The resulting best-fit value of $\sigma=\timeform{3'.1}$ is
reasonable, given the resolution of the SFD map ($\timeform{2'.59}$ in
Gaussian width) and the additional smoothing due to the
$\timeform{2'.37}$ pixelisation and our cloud-in-cell interpolation.
 
\begin{figure}[bt]
\begin{center}
   \FigureFile(70mm,80mm){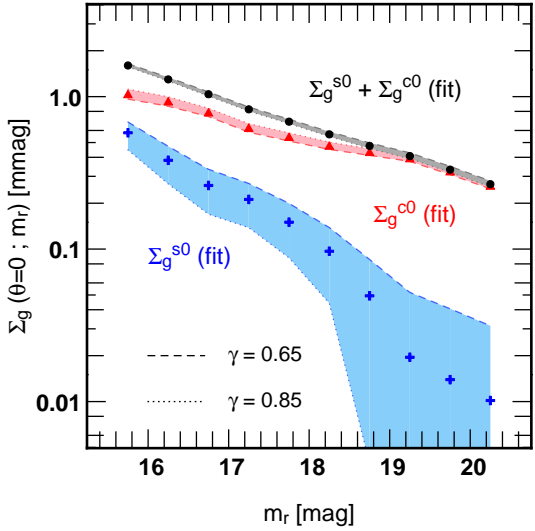}
\end{center}
\caption{Best-fit parameters characterizing the FIR emission of galaxies
 against their $r$-band magnitude.  Crosses, triangles,
 and circles indicate the best-fit value of $\Sigma_{\rm g}^{\rm s0}$
 (fit), $\Sigma_{\rm g}^{\rm c0}$ (fit), and
 $\Sigma_{\rm g}^{\rm s0}$(fit) $+ \Sigma_{\rm g}^{\rm c0}$(fit),
 respectively, assuming $\gamma=0.75$. Shaded
 regions indicate the variance of the best-fit parameters for
 $0.65<\gamma<0.85$. Dashed curve indicates the results for
 $\gamma=0.65$ and dotted curve is for $\gamma=0.85$.  }
 \label{fig:Sigmag-rmag}
\end{figure}

\begin{figure}[bt]
\begin{center}
    \FigureFile(70mm,80mm){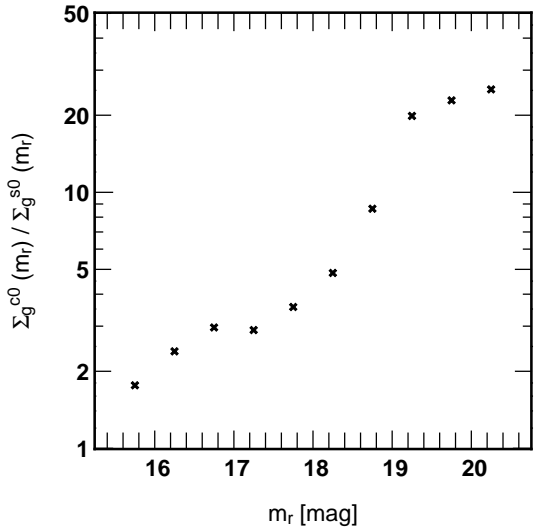}
\end{center}
\caption{Ratio of the clustering term and the central galaxy
 contribution as a function of the $r$-band magnitude of the central galaxy.
}  \label{fig:Sigmag-ratio-rmag}
\end{figure}

The best-fit parameters for $\Sigma_{\rm g}^{\rm s0}(m_r)$ and
$\Sigma_{\rm g}^{\rm c0}(m_r)$ are plotted in Figure
\ref{fig:Sigmag-rmag}.  Even at the central position of the stacked
images, the FIR signals are indeed dominated by the clustering term
$\Sigma_{\rm g}^{\rm c}$ rather than the single galaxy term (see Figure
\ref{fig:Sigmag-ratio-rmag}). Since the value of
$\gamma$ is somewhat uncertain and also depends on magnitudes, $m_r$ and
$m_r^{\prime}$, we performed the same fitting analysis by varying the
value for $0.65 < \gamma < 0.85$, which roughly covers the range of
$\gamma$ for our sample magnitudes.  The dashed lines in Figure
\ref{fig:Sigmag-rmag} show best-fit values for $\gamma = 0.65$, and
dotted lines for $\gamma = 0.85$.  Although the best-fit values of
$\Sigma_{\rm g}^{\rm s0}(m_r)$ are affected by the choice
of $\gamma$, especially for the faint magnitudes, the results for
$\Sigma_{\rm g}^{\rm c0}(m_r)$ hardly change and still dominate the
single term.

\begin{figure}[bt]
\begin{center}
    \FigureFile(70mm,80mm){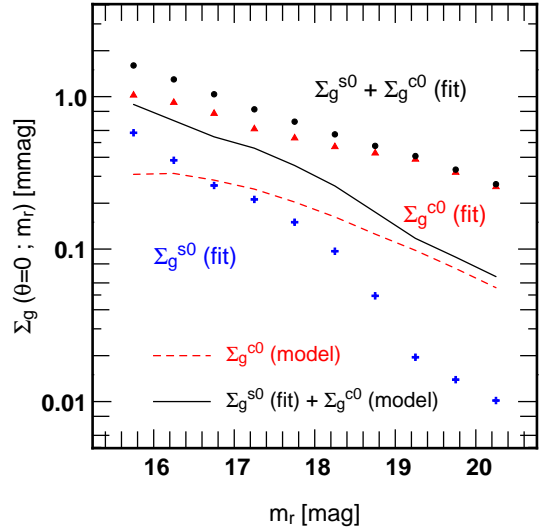}
\end{center}
\caption{The clustering term $\Sigma_{\rm g}^{\rm
c0}$(model) estimated from the SDSS angular cross-correlation is plotted
as a function of the $r$-band magnitude of the central galaxy (dashed
curve).  Solid curve indicates the sum of the single and clustering term
$\Sigma_{\rm g}^{\rm s0}$(fit) $+ \Sigma_{\rm g}^{\rm c0}$(model).  The
symbols are the same as Figure \ref{fig:Sigmag-rmag}, and plotted for
comparison.  } \label{fig:Sigmag-model}
\end{figure}

In reality, we did not anticipate this before performing the radial
fits.  Therefore we directly measure the angular
cross-correlation of SDSS galaxies for comparison, and independently
estimate the clustering term using equation
(\ref{eq:clustering-galaxy-profile}), which we call $\Sigma_{\rm g}^{\rm
c0}$(model). Dashed and Solid curves in Figure \ref{fig:Sigmag-model} indicate
$\Sigma_{\rm g}^{\rm c0}$(model) and $\Sigma_{\rm g}^{\rm
s0}$(fit)$+\Sigma_{\rm g}^{\rm c0}$(model), respectively. We still adopt 
$\Sigma_{\rm g}^{\rm s0}$(fit) (blue crosses), obtained by the profile-fits above, for the 
single term. The results
indicate that resolved SDSS galaxies explain only 25$\sim$40\% of the
clustering term.  This implies that the relation between $r$-band and
FIR luminosity has significant scatter and/or substantial fraction of
FIR sources are not resolved by the SDSS optical selection criteria.
The estimated clustering term exceeds the single
term by more than a factor of a few for faint magnitudes.

\begin{figure}[bt]
\begin{center}
    \FigureFile(70mm,80mm){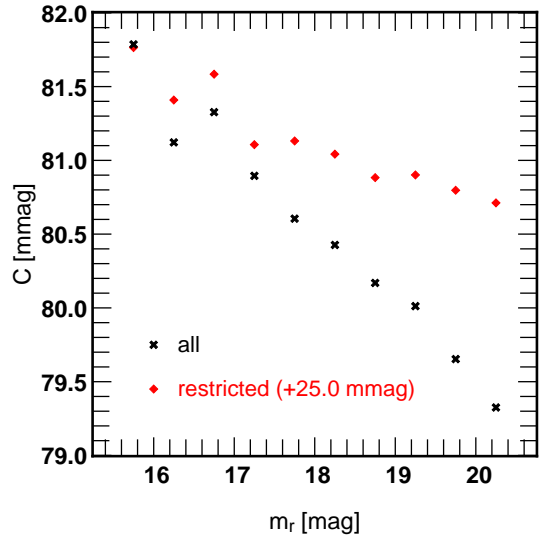}
\end{center}
\caption{The background noise level $C$ against the $r$-band magnitude of
 the central galaxy.
Black crosses indicate the results for all our sample and red ones 
 are restricted in the inner regions shown as yellow lines in Figure \ref{fig:region-SFDmap}.
 }  \label{fig:C-rmag}
\end{figure}

The fitted values of the background noise term $C$ are plotted against
$m_r$ in Figure \ref{fig:C-rmag}. According to our model, they should
not depend on $m_r$, but it is not the case at all.  A systematic
decrease of $C$ against $m_r$ is clearly seen.  We are not yet able to
understand this systematic behavior.  We repeated the same analysis by
selecting those galaxies located in the inner contiguous regions
($160^{\circ}<\alpha<220^{\circ}$, $5^{\circ}<\delta <80^{\circ}$;
see Figure \ref{fig:region-SFDmap}).  The results are
plotted in red crosses after shifting 25 mmag, just for the ease of
visual comparison. The fact that the value of $C$ is also sensitive to
the region of the map and the weaker $m_r$-dependence in this case may
imply that this is due to a subtle selection criterion difference of
SDSS photometric galaxies as a function of the Galactic extinction
itself, including the star-galaxy separation in different area, for
instance. We do not investigate this issue in the present paper since
our main finding is unlikely affected by the position dependent value of
the background noise term.  Actually, the best-fit
values for other quantities are not sensitive to the particular choice
of subregions in the SFD map.

Incidentally the small value of $C$ with respect to the general trend
at $16.0< m_r <16.5$ is the reason why the corresponding profile
in Figure \ref{fig:galaxy-profile} does not follow the systematic trend
of the other profiles.

We note that, if $\sigma$ is treated as a free
parameter separately for different magnitudes, $\sigma(m_r)$, the
resulting fits in the cases of $m_r > 19.0$ return a small negative
value for $\Sigma_{\rm g}^{\rm s0}$.  We found that this is simply due
to the fact the total signal is dominated by the clustering term; the
unambiguous extraction of the single galaxy contribution in those cases
is difficult if we add another degree of freedom in $\sigma$ for each
magnitude bin.  In the case of $m_r<19$, the best-fit values of $\sigma$
are indeed almost independent of $m_r$ and $\sigma \sim \timeform{3'.1}$
as expected from our model assumption.  This is why we constrained
$\sigma$ to be independent of $m_r$ in the actual fitting
procedure. This assumption does not affect the best-fit parameters
except for $\Sigma_{\rm g}^{\rm s0}$ with $m_r>19$.

\section{Implications}

The fitting results presented in the previous section are
model-independent in a sense that it does not assume any {\it a priori}
relation between the $r$-band magnitude and the FIR emission of
galaxies. Therefore the empirical correction of the SFD extinction for
galaxies can be made from Figure \ref{fig:Sigmag-rmag}. We note here
that the correction is not limited to SDSS galaxies at all, and can be
directly applied to future galaxy surveys as long as the $r$-band
magnitude of galaxies is measured. This is because the dominant
clustering term is contributed by galaxies that are not even resolved or
identified in the SDSS galaxy catalog.

Nevertheless it is interesting to consider the underlying connection
between the FIR emission and $r$-band magnitude of galaxies on the basis
of the present result.
Consider a galaxy with intensity profile $I_{\rm 100\mu
  m}(\boldsymbol \theta)$[MJy/sr].  Then its contribution to the $r$-band extinction
  should be
\begin{equation}
\label{eq:I100-dA}
\Sigma_{\rm g}^{\rm s}(\boldsymbol \theta;m_r) = \left[\frac{A_r}{E(B-V)}\right] \times p
 \times I_{\rm 100\mu m}(\boldsymbol \theta),
\end{equation}
where $A_r/E(B-V)$ and $p$ are the conversion factors from the color
excess $E(B-V)$ to the $r$-band extinction and from 100$\mu$m intensity
to $E(B-V)$, and are given as 2.751 and 0.0184, respectively
\citep{SFD98}. 
Integrating equation (\ref{eq:I100-dA}) over $\boldsymbol \theta$
assuming the Gaussian profile, we obtain

\begin{equation}
\label{eq:sigmas0-f100}
2\pi\sigma^2 \Sigma_{\rm g}^{\rm s0}(m_r)
= \left[\frac{A_r}{E(B-V)}\right] \times p 
\times f_{\rm 100\mu m} ~{\rm [MJy]} .
\end{equation}
Finally the 100$\mu$m flux, $f_{\rm 100\mu m}$ is translated to the
100$\mu$m magnitude:
\begin{equation}
m_{\rm 100\mu m}=-2.5\log (f_{\rm 100\mu m}/3.63\times10^{-3}[\rm{MJy}]),
\end{equation}
and equation (\ref{eq:sigmas0-f100}) is rewritten in terms of 
$m_r$ as
\begin{equation}
\label{eq:m100-dA}
\Sigma_{\rm g}^{\rm s0}(m_r) = 36.0 \times 10^{-0.4m_{\rm 100\mu m}} 
\left(\frac{3'.1}{\sigma}\right)^2.
\end{equation}

\begin{figure}[bt]
\begin{center}
    \FigureFile(70mm,80mm){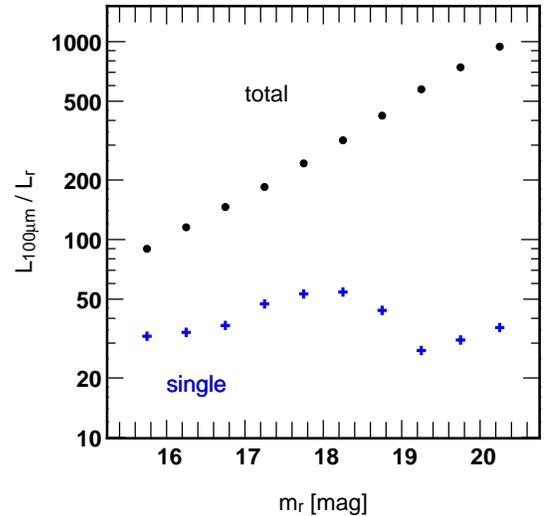}
\end{center}
\caption{Ratio of FIR and $r$-band luminosities as a function of the
 $r$-band magnitude of the central galaxy. Crosses
 indicate the ratio for single galaxy term, while circles include the
 clustering term as well.  }\label{fig:FIR-rband-ratio}
\end{figure}

Since those magnitudes, $m_r$ and $m_{\rm 100\mu m}$, should correspond
to the same galaxy, their difference is equivalent to the ratio of their
absolute luminosities, $L_{\rm 100\mu m}/L_r$. Thus estimated ratios are
plotted in Figure \ref{fig:FIR-rband-ratio}.  The fact that the ratio
for a single galaxy is approximately constant indicates that the
statistically averaged ratio of FIR and optical luminosities of galaxies
are independent of the $r$-band magnitude, which is very reasonable.
For comparison, we plot the ratio for adding the clustering term in
FIR. In this case the integration of $\Sigma_{\rm g}^{\rm c}(\boldsymbol
\theta;m_r)$ over $\boldsymbol \theta$ does not converge because our
assumed value of $\gamma(=0.75)$ is valid only for angular separation
less than $1^\circ$. Thus we evaluate the flux simply by multiplying
$2\pi \sigma^2$ as in the case of the Gaussian profile. Therefore
$L_{\rm 100\mu m}$ includes the contribution of other galaxies, but the
total amplitude is subject to change depending on the more accurate
profile at larger angular scales.  The total ratio follows a clear
single power-law and we believe that the wiggles of the ratio for single
galaxies is not real but comes from the difficulty in separate the
single galaxy contribution from the total signal as mentioned at the end
of the previous section.

\begin{figure}[tb]
\begin{center}
\FigureFile(70mm,80mm){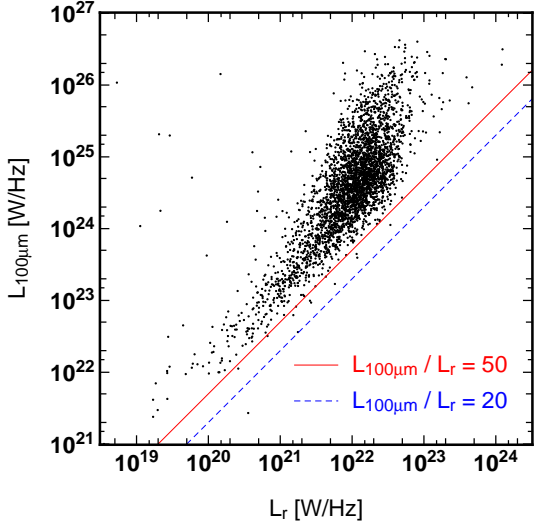}
\end{center}
\caption{Relation between $L_{100\mu\rm{m}}$ and $L_r$ for the IRAS/SDSS
 overlapped galaxies. The solid and dashed lines
 indicate $L_{\rm 100\mu m}/L_r = 20$, and $50$, respectively.}
 \label{fig:nuL-distribution}
\end{figure}

The relation between FIR and optical luminosities of galaxies can be
directly probed from the sample of 3304 galaxies overlapped in the SDSS
and PSCz \citep{Saunders2000}.  We select the brightest SDSS galaxy
locating within 2 arcmin from each PSCz galaxy as its counterpart.  This
procedure enables us to find SDSS optical counterparts for $\sim95\%$ of
the PSCz galaxies that are located in the SDSS survey region.
We apply the K-correction based on the ``K-corrections calculator''
service \citep{Chilingarian2010} for $r$-band, and extrapolate the FIR
spectral energy densities using second-order polynomials determined from
25, 60 $\mu\rm{m}$ flux for 100 $\mu \rm{m}$ \citep{Takeuchi2003}.
Figure \ref{fig:nuL-distribution} is a scatter plot of
$L_{100\mu\rm{m}}$ (PSCz) and $L_r$ (SDSS) for
the PSCz--SDSS overlapped sample of galaxies.
Figure \ref{fig:nuL-distribution} indicates an approximate linear
relation between $L_{100\mu\rm{m}}$ and $L_r$ albeit with considerable
scatters. The solid lines correspond to $L_{\rm 100\mu m}/L_r=20$ and 50
for reference.  Thus the approximate linear relation implied from Figure
\ref{fig:FIR-rband-ratio} is largely consistent with Figure
\ref{fig:nuL-distribution} since the IRAS galaxies may be preferentially
biased toward the FIR brighter ones than average.  In turn, this
confirms that our interpretation that $\Sigma_{\rm g}^{\rm s}$
represents the contribution of a single galaxy.

\section{Stacking analysis of SDSS DR6 quasars \label{sec:quasars}}

Y07 did not find any definite anomaly between the quasar surface
densities and $A_{\rm SFD}$ beyond the statistical errors. Nevertheless
it is interesting to repeat the stacking analysis to SDSS quasar sample
as well. Indeed as we show below, we found that a weaker but similar
pattern of the enhanced extinction around stacked quasar images.

\begin{figure*}[tb]
\begin{center}
     \FigureFile(45mm,45mm){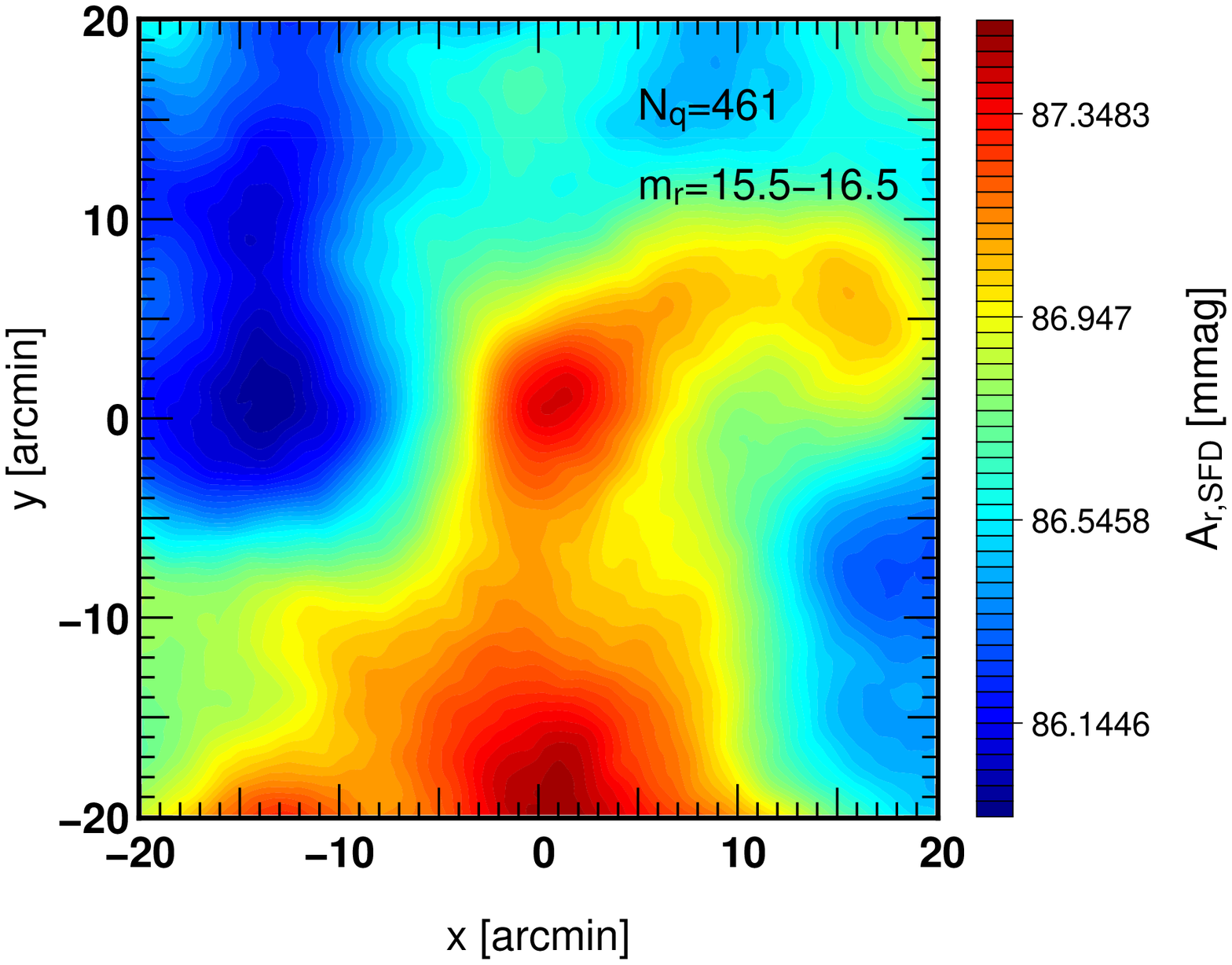}
     \FigureFile(45mm,45mm){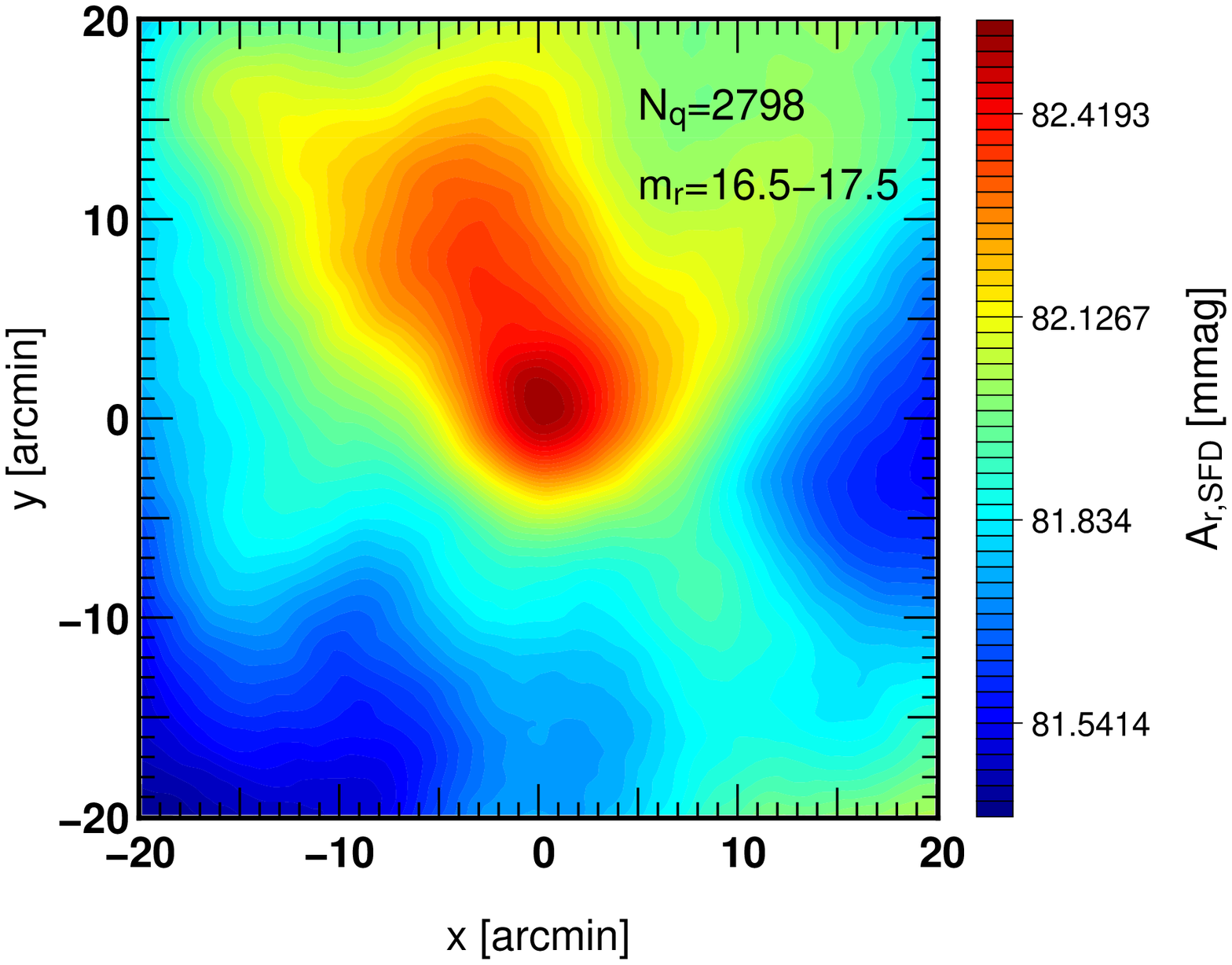}
     \FigureFile(45mm,45mm){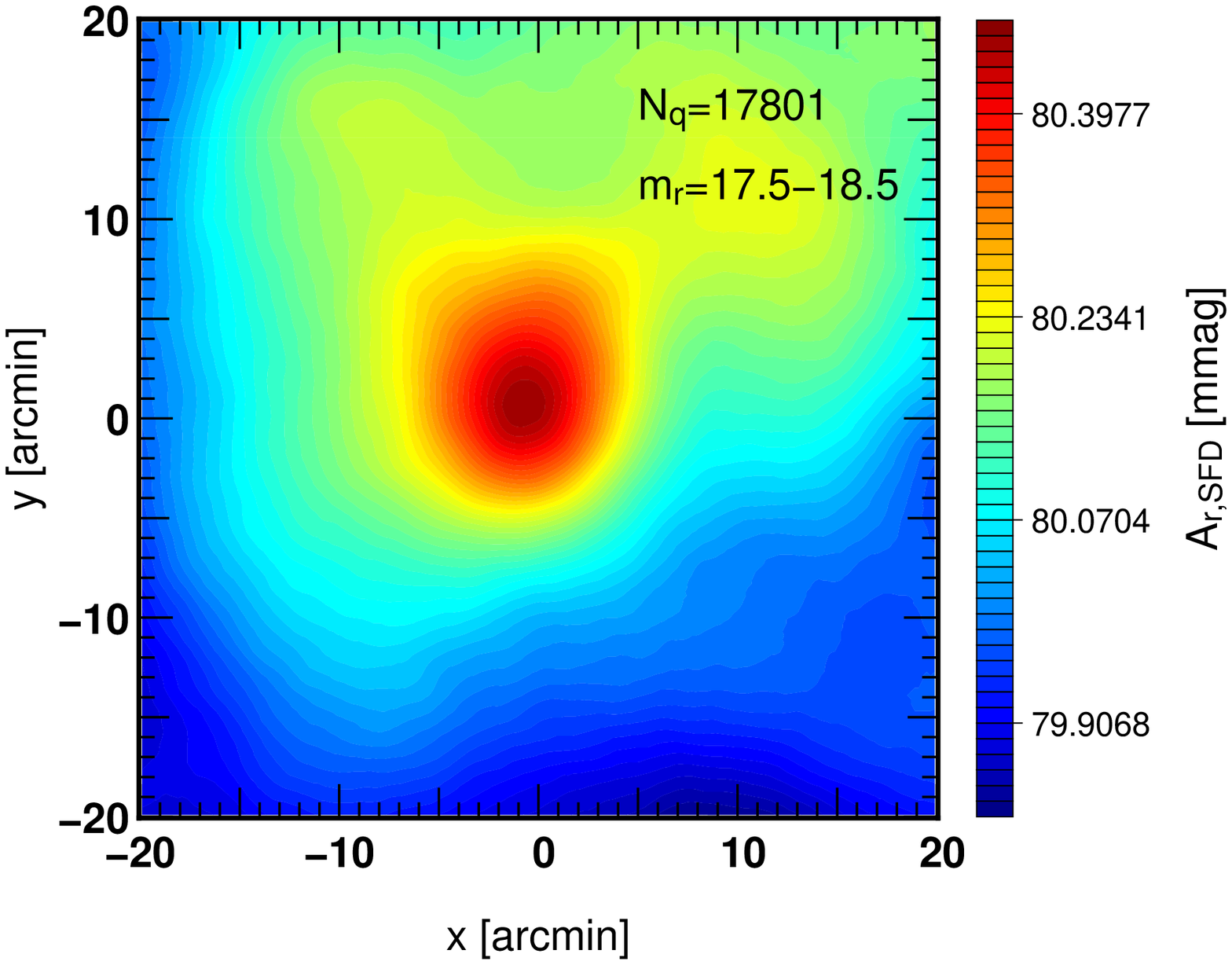}
     \FigureFile(45mm,45mm){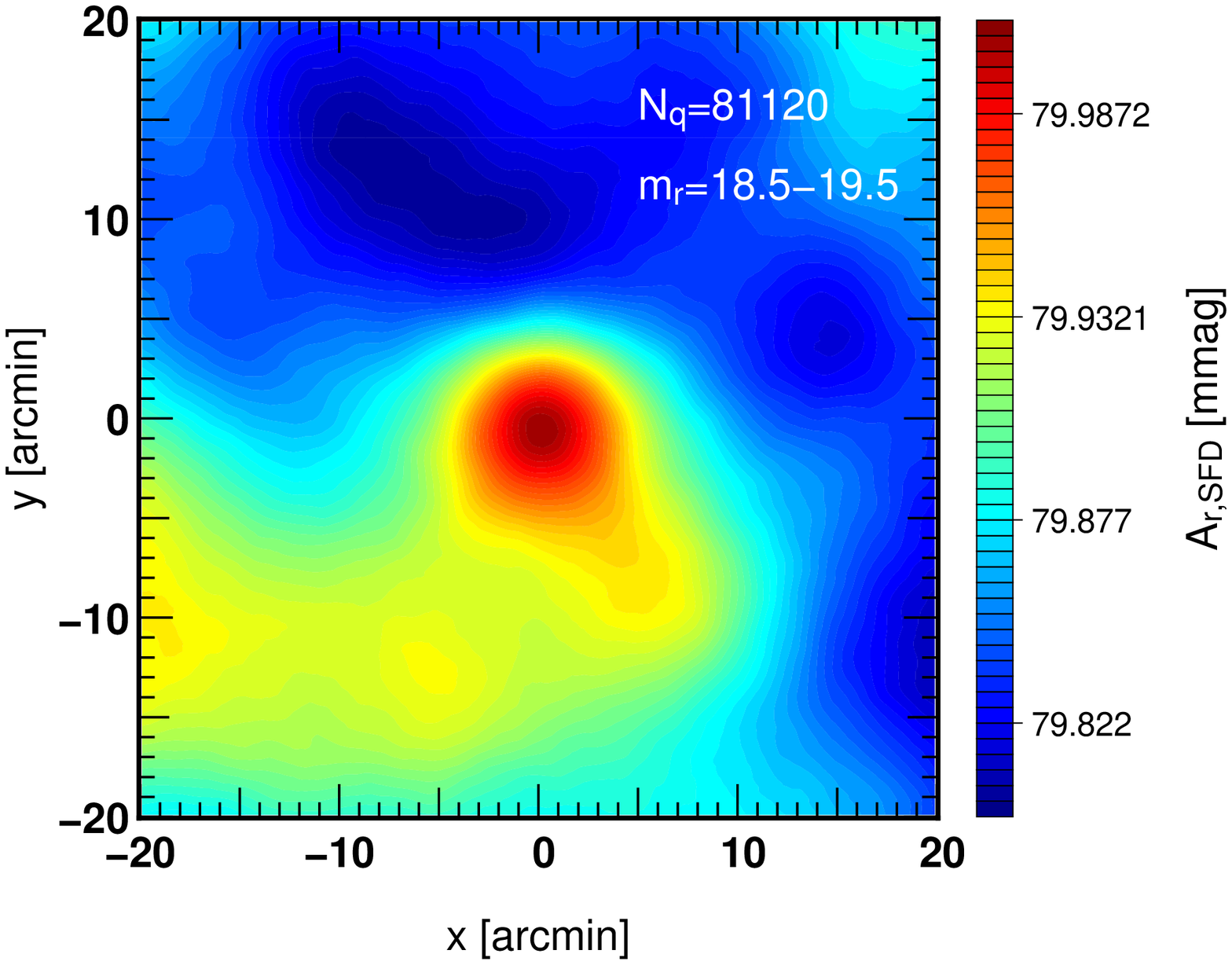}
     \FigureFile(45mm,45mm){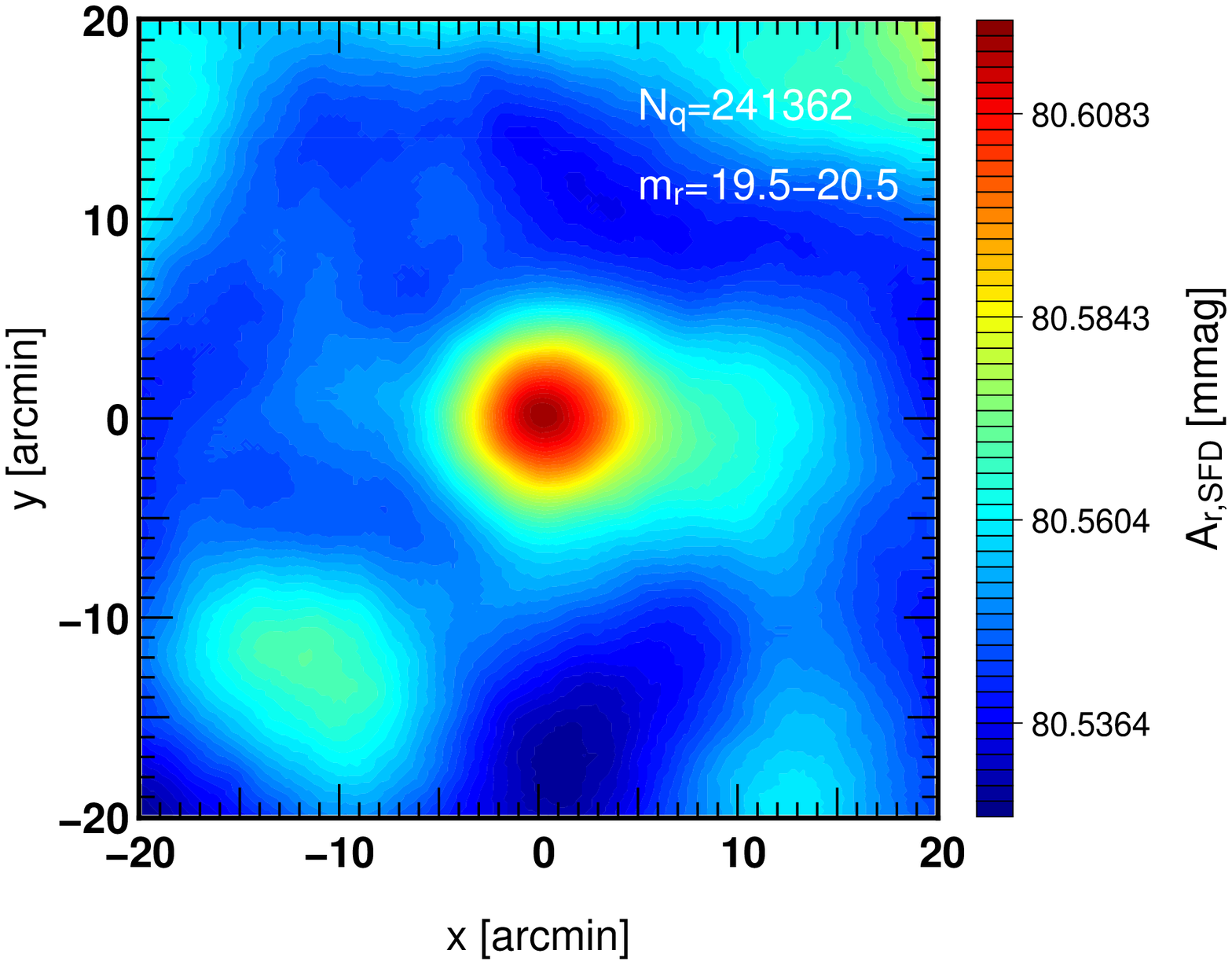}
 
\end{center}
\caption{Stacked images of the SFD map for $40'\times40'$ centered at
SDSS quasars of different $r$-band magnitudes
$(m_r=15.5\sim20.5~\rm{mag})$ in 1.0 magnitude bin. The magnitude range
and the number of quasars in the range are denoted in each panel.  }
\label{fig:quasar-diff-mag}
\end{figure*}

\begin{figure*}[bt]
\begin{center}
 \FigureFile(45mm,45mm){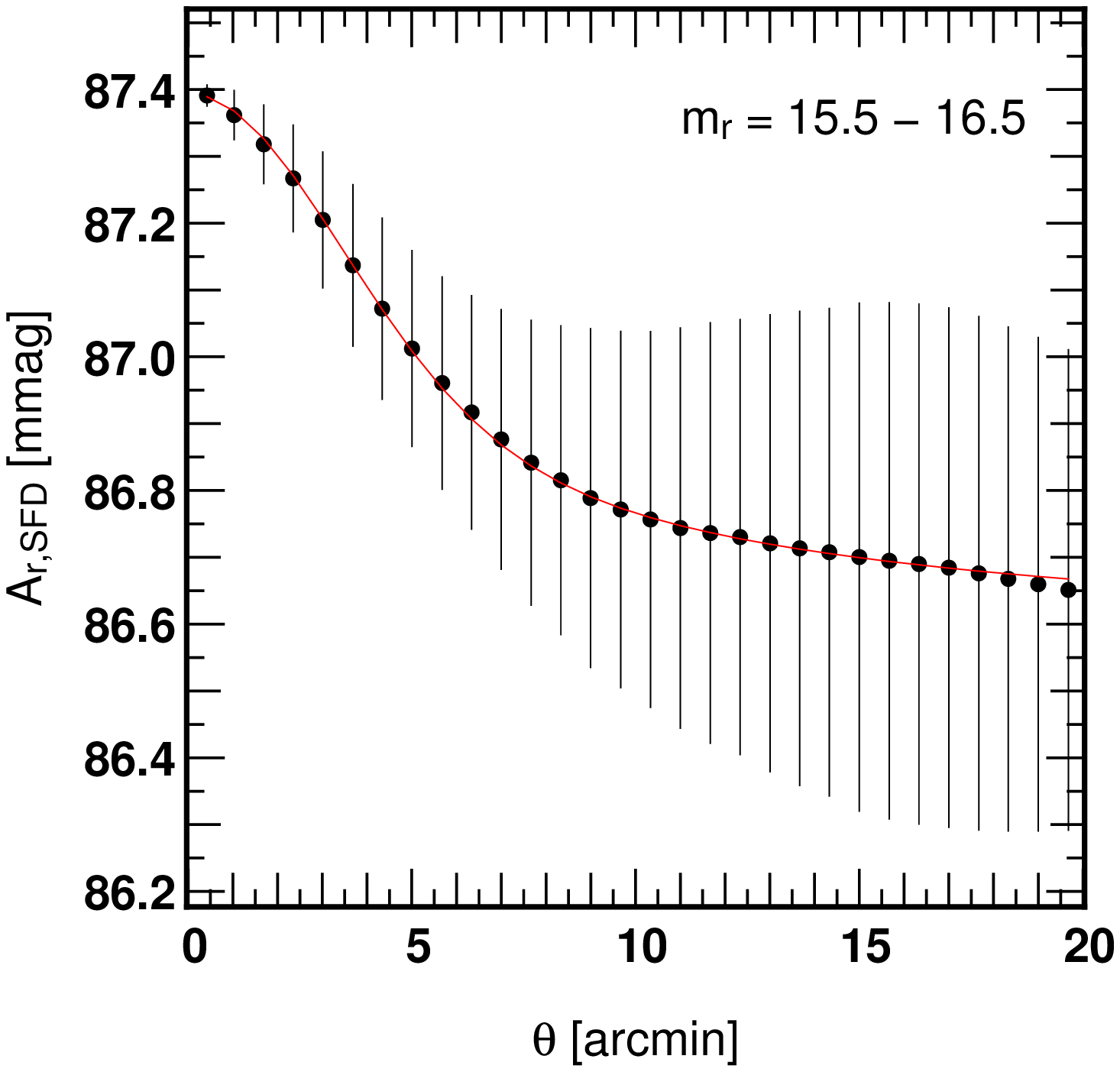}
        \FigureFile(45mm,45mm){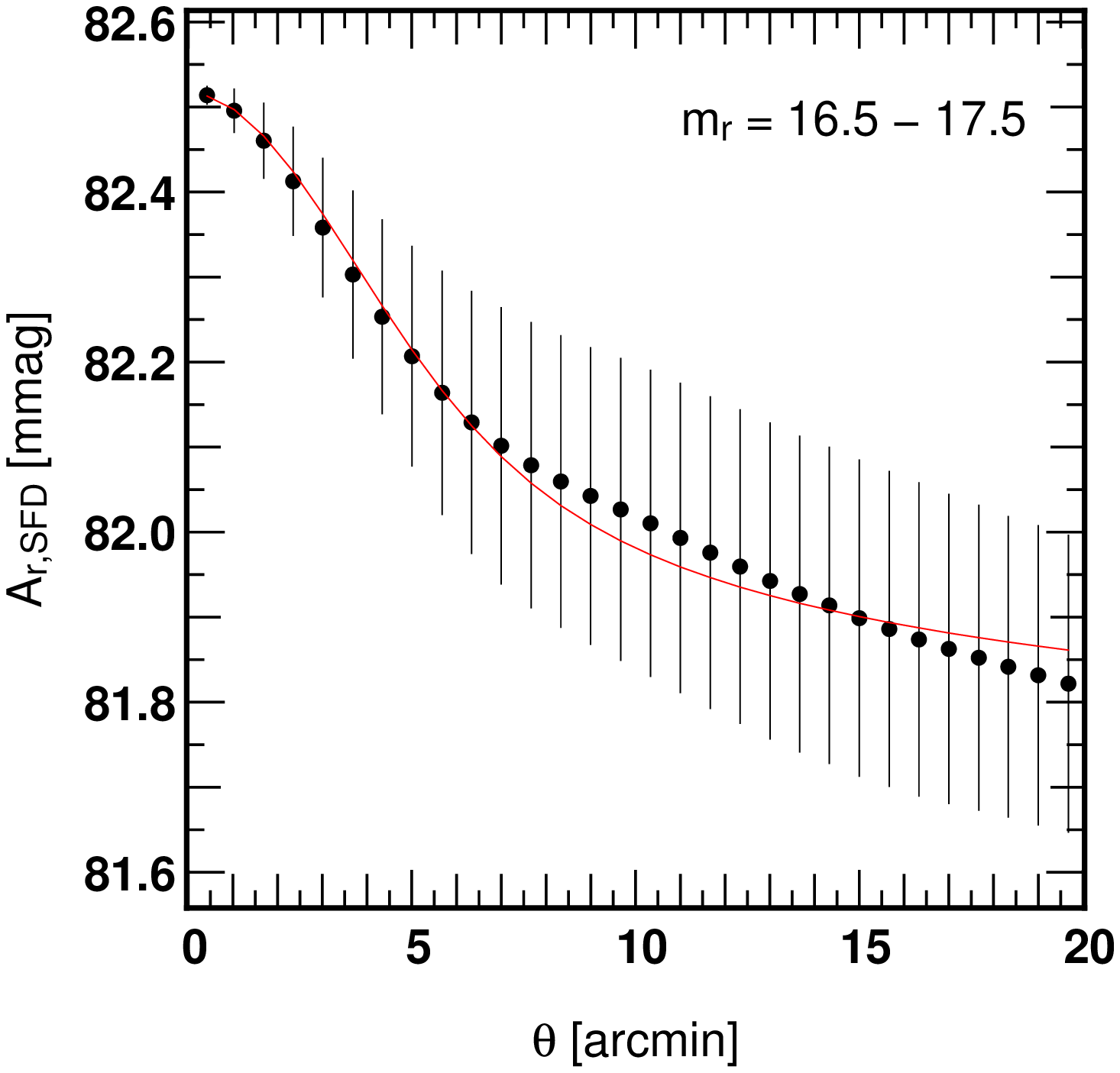}
        \FigureFile(45mm,45mm){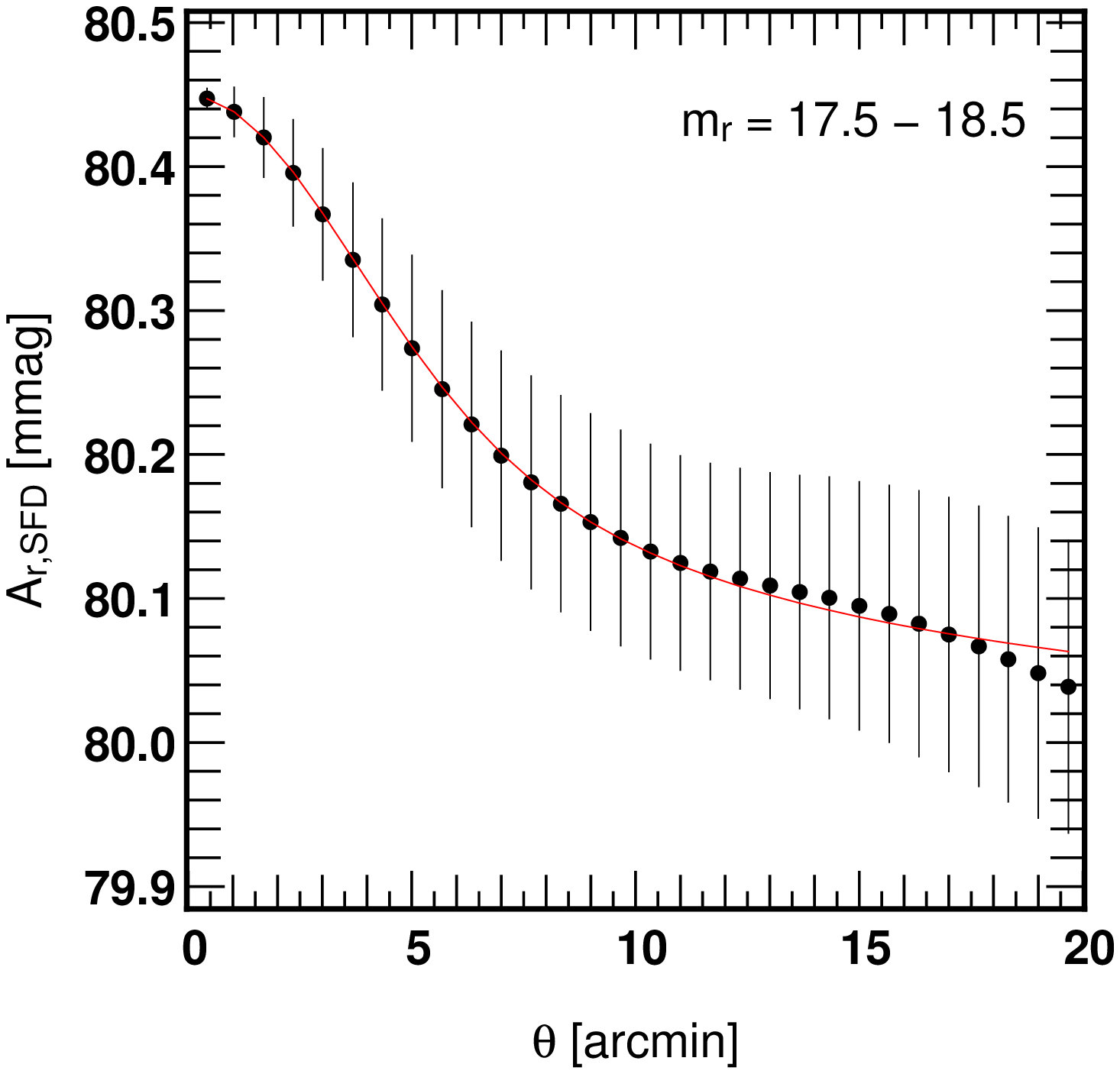}
        \FigureFile(45mm,45mm){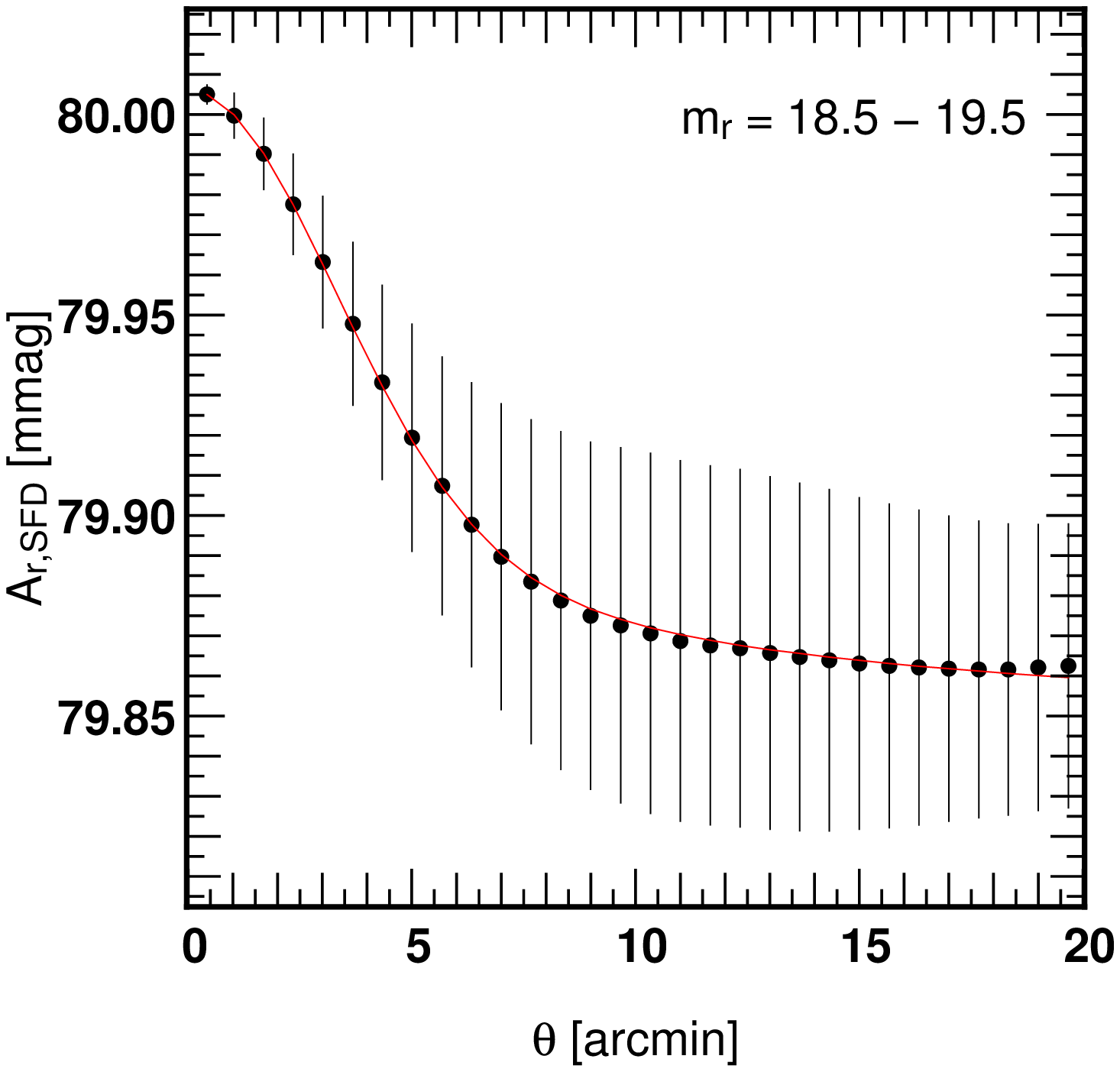}
        \FigureFile(45mm,45mm){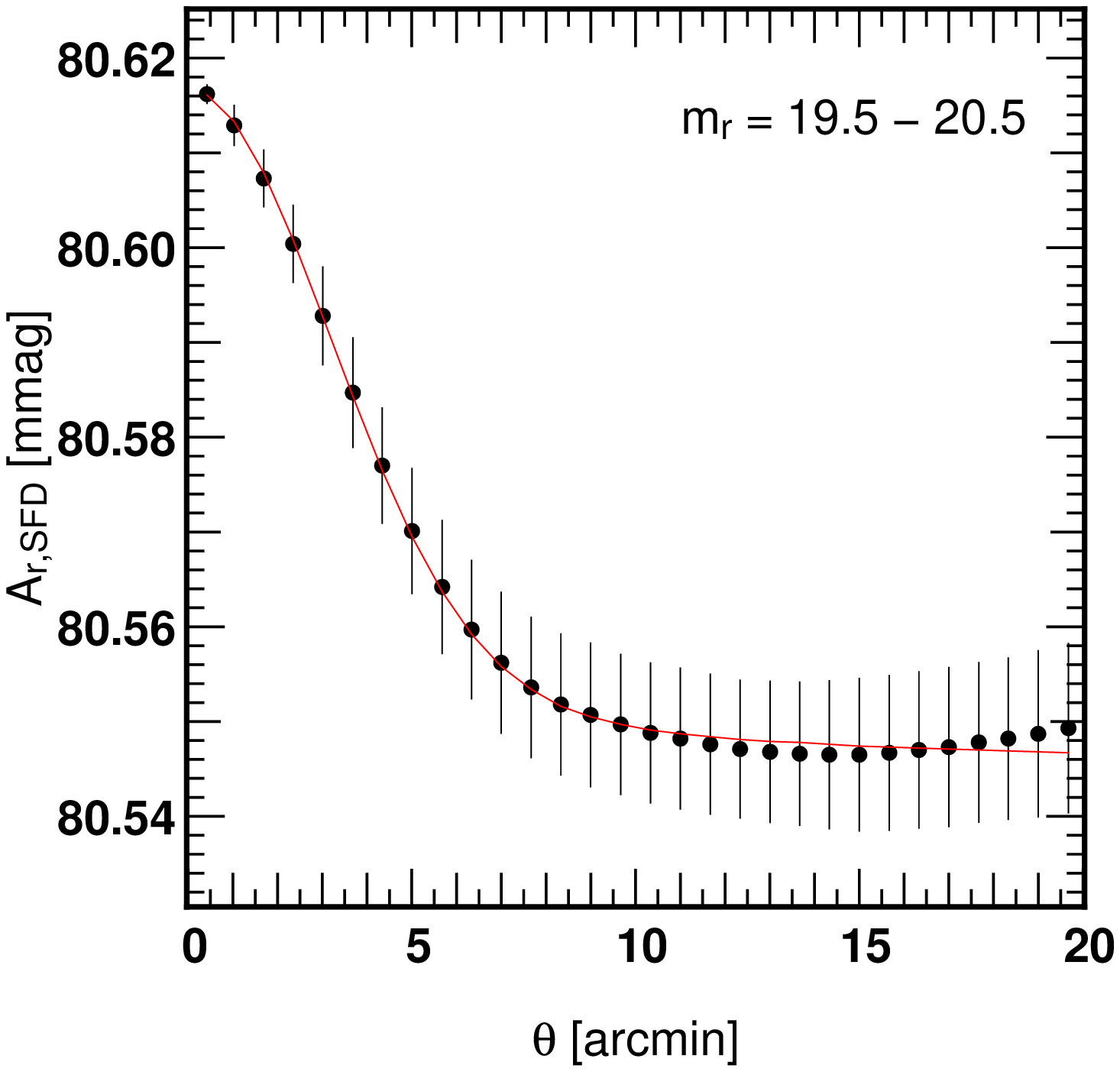}
\end{center}
\caption{Radial profiles of stacked quasar images corresponding to
Fig. \ref{fig:quasar-diff-mag}. Solid curves indicate
the best-fit model of equation (\ref{eq:average-galaxy-profile}),
(\ref{eq:single-galaxy-profile}), and
(\ref{eq:clustering-galaxy-profile2}).
}  \label{fig:quasar-profile}
\end{figure*}

For this purpose, we use the SDSS DR6 photometric quasar sample
\citep{Richards2009a, Richards2009b}.  The analysis method is basically
identical to that performed in subsection \ref{subsec:galaxies} except
that we have to use the larger magnitude bins ($\Delta m_r=1.0$)
due to the limited number of the quasars as well as the weaker signature of
the effect. The stacked images are plotted in Figure
\ref{fig:quasar-diff-mag}. As in the case of galaxies, we fit the radial
profile to equations (\ref{eq:average-galaxy-profile}),
(\ref{eq:single-galaxy-profile}), and (\ref{eq:clustering-galaxy-profile2}) assuming that $\sigma$ is 
independent of $m_r$.
We find the best-fit value of $\sigma=\timeform{3'.13}$, which is similar to that
 for galaxy radial profile.
 The radial profiles and the best-fit curves are plotted in Figure
\ref{fig:quasar-profile}.

Unlike galaxies, the profiles are not completely circular, and also the
best-fit parameters do not exhibit regular behavior as a function of
$m_r$. Part of the behavior may be due to the contamination of
non-quasars objects in the photometric quasar sample. Therefore it would be
better to repeat the analysis for the spectroscopic quasar sample, which
we plan to do in due course.  Nevertheless the results indicate a clear
signal around the center of all the tacked images. If we look at Figure
9 in Y07 carefully, a very weak anomaly may be recognized for
photometric quasars as well. This would be consistent with our current
finding of the FIR emission around those quasars in the SFD map.

It is interesting to ask if the detected FIR emission around quasars
originated from (1) quasars themselves, (2) their host galaxies, (3)
neighbor galaxies due to the quasar-galaxy and/or quasar host galaxy-galaxy
 correlation, and/or (4) some other effects (lensing, for
instance). Observational studies of quasar V-band luminosity $M_{\rm q}$
and that of the host galaxy $M_{\rm g}$ imply a very weak correlation,
at most, with significant scatters. Typically a quasar is one or two
magnitudes brighter than its host galaxy
\citep{Hamilton08,Letawe10}. Given those combined with the results for
galaxies discussed in \S 3, the signal may be ascribed to the
possibility (3). This deserves further investigation, and we are
currently working on the analysis using the WISE \citep{Wright2010} and IRAS \citep{Wheelock1994} data to
explore the quasar-galaxy correlation in FIR.

\begin{figure}[bt]
\begin{center}
\FigureFile(70mm,80mm){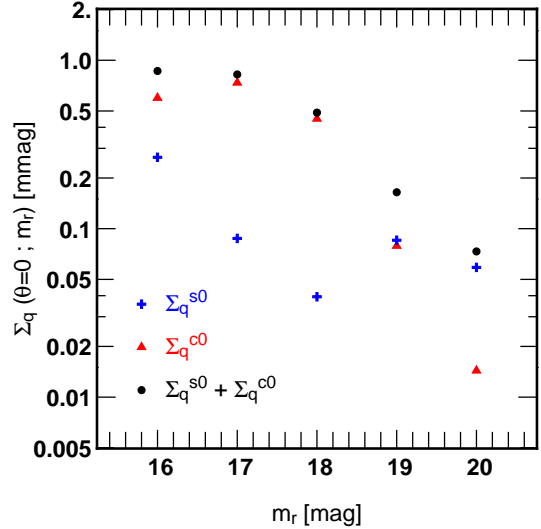}
\end{center}
\caption{Best-fit parameters characterizing the FIR emission of quasars
against their $r$-band magnitude.  }  \label{fig:Sigmag-rmag-qso}
\end{figure}

\begin{figure}[bt]
\begin{center}
\FigureFile(70mm,80mm){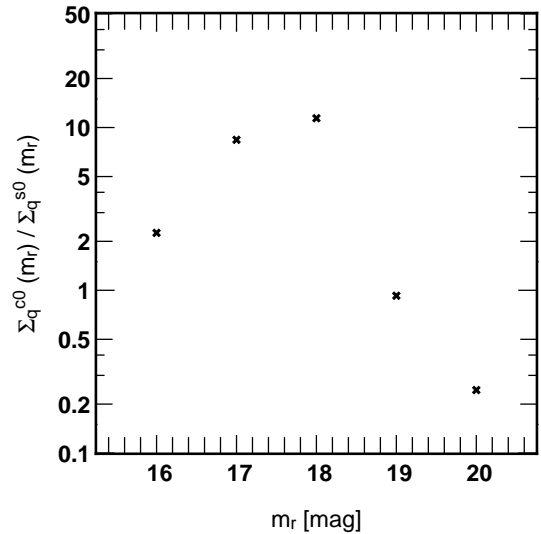}
\end{center}
\caption{Ratio of the clustering term and the central quasar
 contribution as a function of the $r$-band magnitude of the central
 quasar.  }  \label{fig:Sigmag-ratio-rmag-qso}
\end{figure}

\section{Conclusions}

We have detected the small but systematic contamination in the SFD
Galactic extinction map due to FIR emission from galaxies. The amount of
the contamination of the order of mmag is very small and may be
negligible for most astronomical purposes. Nevertheless, the
contamination is intrinsically correlated with the large-scale structure
of the universe traced by galaxies, and therefore should be kept in mind
in precision cosmological measurements. Indeed even such a small
contamination leads to a relatively large anomaly of the galaxy surface
number density as a function of $A_{\rm SFD}$ discovered by
\citet{Yahata2007}.

\begin{figure}[bt]
\begin{center}
    \FigureFile(70mm,80mm){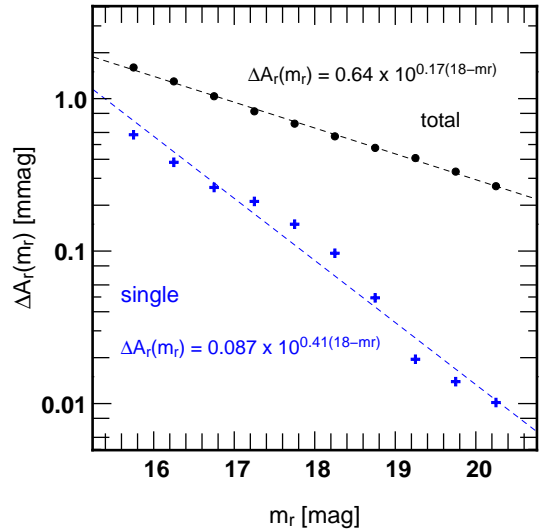}
\end{center}
\caption{$\Delta A_r$ as a function of $r$-band magnitude of the central
 galaxy. Crosses indicate the single galaxy term,
 while circles include the clustering term as well.}
 \label{fig:deltaa-rmag}
\end{figure}

Our major result is that a galaxy of $r$-band magnitude $m_r$ has an
additional {\it contribution} to the SFD Galactic extinction by an
amount of 
\begin{equation}
\label{eq:dA-mr-gal-single}
\Delta A_r(m_r) = 0.087 \times 10^{0.41(18-m_r)}~{\rm [m mag]},
\end{equation}
due to the FIR emission from itself (single term), and

\begin{equation}
\label{eq:dA-mr-gal-total}
\Delta A_r(m_r) = 0.64 \times 10^{0.17(18-m_r)}~{\rm [m mag]},
\end{equation}
including the contribution from nearby galaxies
(clustering term: Figure \ref{fig:deltaa-rmag}).
Note that since the SFD determination of conversion
factor $p$ has statistical and systematic uncertainties of approximately
8\%, equation (\ref{eq:dA-mr-gal-single}) and (\ref{eq:dA-mr-gal-total})
would have the similar level of uncertainties.

One possible method to correct the SFD map is to subtract all the contributions
 of identified galaxies; the corrected extinction at position $\boldsymbol \theta$ is obtained as
\begin{equation}
\label{eq:dA-mr-pix}
A'_r({\boldsymbol \theta}) = A_{r,\rm SFD}({\boldsymbol \theta}) 
- \sum_{j} \Sigma_{\rm g}^{\rm s} ({\boldsymbol \theta_j - \boldsymbol \theta};m_r^j),
\end{equation}
where ${\boldsymbol \theta_j}$ is the position of the $j$-th galaxy with its $r$-band
magnitude of $m_r^j$. Note that this correction ignores contributions
of other than the identified galaxies; if the SDSS DR7 photometric galaxy
catalog is used, only contributions from those galaxies listed in the
catalog are corrected. We could use $\Sigma_{\rm g}^{\rm tot} 
({\boldsymbol \theta_j - \boldsymbol \theta};m_r^j)$ in equation
(\ref{eq:dA-mr-pix}), including the clustering term $\Sigma_{\rm g}^{\rm c}$.
In that case, however, it would over-count the true contribution
especially when the pixel has many nearby galaxies.  In turn it is
interesting to examine the extent of which the anomaly detected by Y07
is explained by the SDSS DR7 galaxies alone, by applying the correction on
the basis of equation (\ref{eq:dA-mr-pix}).

Most likely improved correction formulae than equation
(\ref{eq:dA-mr-gal-single}) and (\ref{eq:dA-mr-gal-total}) may be obtained by further dividing galaxy samples
according to their colors, fully exploiting the multi-band photometries
of SDSS DR7. Also similar stacking analyses of various current/future
sky maps in different wavelengths, e.g., WISE, IRAS, AKARI \citep{Murakami2007},
 WMAP \citep{Bennett2003}, Planck \citep{Tauber2010} among
others, are interesting to make sure of their validity and/or put
constraints on the possible contamination.

We would like to emphasize also that the present methodology is useful
to derive statistical relations among galaxy parameters in different
bands and to infer the angular cross-correlation of galaxies, much
beyond the magnitude limit for individual galaxy surveys. The results
for those directions will be presented elsewhere.

\bigskip 

We thank Issha Kayo, Shirley Ho, Teruyuki Hirano,
Atsushi Taruya, Toru Yamada, Brice M{\'e}nard, Takashi
Onaka, Hideyuki Izumiura, and Takao Nakagawa for useful discussions.
We also thank an anonymous referee for many valuable
comments.  This work is supported in part from the Grant-in-Aid
No. 20340041 by the Japan Society for the Promotion of Science.
Y.S. and T.K. gratefully acknowledge supports from the Global Scholars Program of
Princeton University and from Global Center for Excellence for Physical Science Frontier 
at the University of Tokyo, respectively.

Funding for the SDSS and SDSS-II has been provided by the Alfred P.
Sloan Foundation, the Participating Institutions, the National Science
Foundation, the U.S. Department of Energy, the National Aeronautics and
Space Administration, the Japanese Monbukagakusho, the Max Planck
Society, and the Higher Education Funding Council for England.  The SDSS
Web Site is http://www.sdss.org/.

The SDSS and SDSS-II are managed by the Astrophysical Research
Consortium for the Participating Institutions. The Participating
Institutions are the American Museum of Natural History, Astrophysical
Institute Potsdam, University of Basel, Cambridge University, Case
Western Reserve University, University of Chicago, Drexel University,
Fermilab, the Institute for Advanced Study, the Japan Participation
Group, Johns Hopkins University, the Joint Institute for Nuclear
Astrophysics, the Kavli Institute for Particle Astrophysics and
Cosmology, the Korean Scientist Group, the Chinese Academy of Sciences
(LAMOST), Los Alamos National Laboratory, the Max-Planck-Institute for
Astronomy (MPIA), the Max-Planck-Institute for Astrophysics (MPA), New
Mexico State University, Ohio State University, University of
Pittsburgh, University of Portsmouth, Princeton University, the United
States Naval Observatory, and the University of Washington.

\appendix

\section*{IRAS Point Spread Function \label{sec:PSF}}

In subsection \ref{subsec:galaxies}, we decompose the
stacked radial profiles into single and clustering terms, assuming the
Gaussian PSF.  Since the IRAS PSF is known to be very complex, we need
to check the validity of this assumption.  For the purpose of
determining the PSF directly, we perform similar stacking analysis with
SDSS stars.

\begin{figure}[bt]
\begin{center}
\FigureFile(70mm,80mm){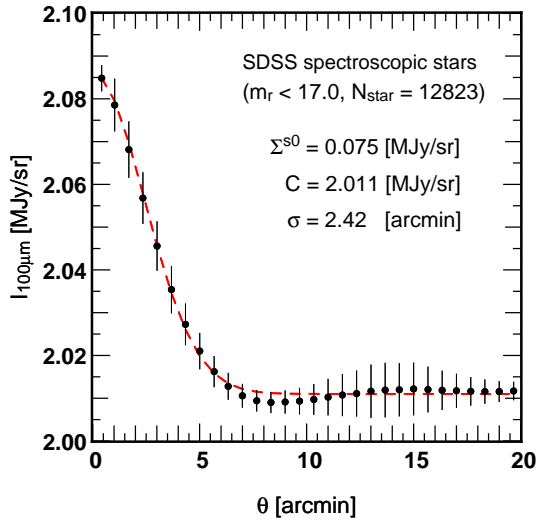}
\end{center}
\caption{Radial profiles of stacked star images.
The dashed curve indicates the best-fit Gaussian profile.} 
\label{fig:star-profile}
\end{figure}

More specifically, we select spectroscopic stars
brighter than $m_r = 17.0$ (12823 stars in total) from the SDSS DR7
catalog.  We first stack the SFD map centered on those SDSS stars as we
did for SDSS galaxies and quasars, but we find no significant signature.
This is mainly because the bright point sources in the IRAS catalogue
are already removed from the SFD map.  Therefore we go back to the
original ISSA 100$\mu m$ diffuse map, and perform the stacking
analysis.  The resulting stacked average radial profile of the SDSS
stars is shown in Figure \ref{fig:star-profile}. The data points are
well approximated by a single Gaussian.  The best-fit Gaussian shown in
the dashed curve has a width of $\sigma = \timeform{2'.42}$, which is
slightly smaller than that found in subsection
\ref{subsec:galaxies}. This is understood because since the SFD map is
constructed by further smoothing the original ISSA map.  Thus we
conclude that our assumption of the Gaussian PSF is valid.

\end{document}